%% file: arxiv_main.tex
\documentclass[11pt]{article}
\usepackage[T1]{fontenc}
\usepackage[utf8]{inputenc}
\usepackage[margin=1in]{geometry}
\usepackage{graphicx}
\usepackage{amsmath,amssymb,bm}
\usepackage{booktabs}
\usepackage{multirow}
\usepackage{array}
\usepackage[numbers,sort&compress]{natbib}
\usepackage[hidelinks]{hyperref}
\usepackage{microtype}
\usepackage{caption}
\usepackage{float}
\graphicspath{{./Figures/}}
\newcolumntype{C}[1]{>{\centering\arraybackslash}m{#1}}

\begin{document}

\title{Diagnosing Latent Energy Decomposition in Machine-Learning Interatomic Potentials via Interacting Quantum Atoms}
\author{Kohei Shimamura\thanks{Corresponding author: \href{mailto:shimamura@kumamoto-u.ac.jp}{shimamura@kumamoto-u.ac.jp}}\\
\small Department of Physics, Kumamoto University, Kumamoto, Japan
\and
Ken-ichi Nomura\\
\small Collaboratory for Advanced Computing and Simulations,\\[-0.2em]
\small University of Southern California, Los Angeles, California, USA}
\date{}

\maketitle

\begin{abstract}
Machine-learning interatomic potentials (MLIPs) can reproduce potential energies and forces accurately, but their internal energy allocation is often difficult to interpret. E3D-IQA is introduced as a diagnostic framework connecting the latent edge-energy representation of an Allegro-type MLIP with Interacting Quantum Atoms (IQA) energy decomposition. The Allegro edge-energy path is retained as a latent pair contribution, while a node-energy path is trained against the IQA intra-atomic energy. IQA interatomic energies are not direct training targets; instead, the learned edge energies are evaluated after training against the IQA pair terms. Tests on H/C/N/O organic reaction structures show that intra-atomic supervision is essential: energy and force training alone does not recover an IQA-like one-body/two-body allocation. With intra-atomic supervision, node energies reproduce IQA intra-atomic terms, and latent edge energies show meaningful correspondence with IQA interatomic terms. Residual errors are concentrated in positive or weak pair interactions, exposing internal allocation failures that remain hidden in total-energy and force metrics. Adding structures labeled only with energies and forces improves transfer to larger molecules and reduces decomposition errors. E3D-IQA therefore provides a route for diagnosing and guiding latent scientific representations using partially labeled quantum-chemical datasets.
\end{abstract}


\section{\label{sec:intro}Introduction}
Machine-learning interatomic potentials (MLIPs) have become an essential tool for molecular and materials simulations because they can approximate first-principles potential-energy surfaces at a much lower computational cost~\cite{Behler_2007,Musil_2021}. Modern equivariant architectures, including NequIP, Allegro, MACE, and related models, achieve high accuracy by constructing expressive many-body representations of local atomic environments~\cite{Batzner_2022,Musaelian_2023,Nomura_2025,Batatia_2023, Batatia_2025}. At the same time, the nonlinear latent representations that make these models accurate also make their internal energy decomposition difficult to interpret~\cite{Schutt_2017,Oviedo_2022}. In most applications, the model is judged by the total potential energy and the total force on each atom, whereas the chemically meaningful allocation of the energy among atoms and atom pairs remains hidden.

This issue is closely related to the non-uniqueness of atom-wise or pair-wise energy assignments in first-principles calculations. A total molecular energy is well defined, but its decomposition into atomic or interatomic contributions is not unique unless an additional theoretical prescription is introduced~\cite{Schutt_2017,Kjeldal_2023}. Consequently, ordinary MLIPs are trained mainly against total energies and forces, and when a model makes an error, it is difficult to identify which atom, bond, or interaction is responsible for the failure. This limits the use of MLIPs not only as fast predictors, but also as tools for chemical analysis.

Recent work suggests that chemically meaningful information can nevertheless emerge in latent representations of MLIPs~\cite{edamadaka_2025}. In particular, emergent energy decomposition (E3D) analysis used the edge-energy structure of Allegro to monitor a latent pair-energy contribution~\cite{Hattori_2026}. In an Allegro-type model, the cohesive energy can be written as
\begin{eqnarray}
E_{\mathrm{coh}} = \sum_i \sum_{j>i} D_{ij},
\label{eq:E3D},
\end{eqnarray}
where $D_{ij}$ is not a fixed classical pair potential, but an edge contribution generated from a many-body local environment. The E3D study showed that this latent edge energy can exhibit bond-dissociation-energy-like behavior even without direct supervision of pairwise energies. This result motivated the question addressed here: can a quantum-chemical energy decomposition be used to diagnose, and partially guide, such latent MLIP energy representations?

To answer this question, we introduce the Interacting Quantum Atoms (IQA) energy decomposition into the E3D framework. IQA is based on the quantum theory of atoms in molecules (QTAIM)~\cite{Bader_1985} and partitions the total energy into intra-atomic and interatomic contributions~\cite{Blanco_2005,Guevara_2020}. In the present notation,
\begin{eqnarray}
E = \sum_i E_{\mathrm{intra}}(i)
  + \sum_i \sum_{j>i} E_{\mathrm{inter}}(i,j).
\label{eq:IQA}
\end{eqnarray}

IQA is particularly informative for interaction analysis because its energy partition is defined in real space by QTAIM atomic basins and does not require atom pairs to be designated as chemical bonds in advance. The intra-atomic term $E_{\mathrm{intra}}$ measures the energetic deformation or strain of an atom relative to its isolated state, whereas $E_{\mathrm{inter}}$ quantifies pairwise interactions between both bonded and nonbonded atoms, including stabilizing and destabilizing contributions. These terms can be further resolved, in principle, into kinetic, electrostatic, exchange, and correlation components. IQA can therefore address not only how strongly two atoms interact, but also which physical contributions underlie that interaction, making it useful for analyzing covalent bonding, noncovalent interactions, atomic strain, and changes in bonding along chemical reactions~\cite{Blanco_2005,Guevara_2020,Symons_2019,JaraCortes_2021}.

This detailed decomposition comes at substantial computational cost~\cite{Guevara_2020}. In addition to the underlying electronic-structure calculation, IQA requires the construction of QTAIM basins and numerical integration of atomic and interatomic energy contributions, commonly through wavefunction-analysis programs such as AIMAll~\cite{Keith_2019}. Repeating this analysis for the large numbers of structures encountered in reaction datasets or molecular-dynamics trajectories is therefore considerably more demanding than evaluating only reference energies and forces.

In this study, we develop E3D-IQA to bridge the detailed but expensive IQA analysis and the rapid inference of an MLIP. E3D-IQA adds an IQA-inspired node-energy path to Allegro while retaining the original edge-energy path. The node energy is trained against $E_{\mathrm{intra}}$, whereas the edge energy $D_{ij}$ is not directly trained against $E_{\mathrm{inter}}$. Once trained on a limited set of IQA-labeled structures, the model can provide approximate IQA-informed node and pair diagnostics without repeating the Psi4/AIMAll workflow for every new structure. E3D-IQA is not intended here as an exact replacement for IQA; rather, it is a scalable framework for transferring part of the information contained in expensive IQA labels to larger structural datasets. The design also follows the E3D observation that meaningful edge energies may emerge without direct pair supervision, while avoiding the instability that could arise from simultaneously constraining total energy, forces, one-body terms, and two-body terms. In addition, it accommodates the practical mismatch between IQA $E_{\mathrm{inter}}$, which is defined for all atom pairs, and MLIP edge energies, which exist only within a finite cutoff.

The goal of this work is to diagnose latent energy decomposition in MLIPs through an IQA-based quantum-chemical reference and to test whether partial interpretability labels can guide the internal representation toward a physically meaningful decomposition. We evaluate E3D-IQA on organic reaction structures containing H, C, N, and O from the Transition1x dataset~\cite{Schreiner_2022}. We show that intra-atomic IQA supervision is necessary for obtaining an IQA-like decomposition, that latent pair energies can correlate with IQA interatomic energies even without direct pair supervision, and that residual errors identify concrete directions for model improvement. We further examine whether conventional energy/force data without IQA labels can extend the learned decomposition to a broader structural domain. This perspective complements existing approaches that use atomistic neural networks for energy decomposition or QTAIM-related learning~\cite{Anstine_2025,Chun_2025,Gallegos_2022,Kjeldal_2023,Takamoto_2022}.

Energy-resolved latent representations may also be useful for reactive molecular dynamics analysis. Existing trajectory-analysis tools such as ChemTraYzer, ReacNetGenerator, ReaxANA, and ChemXDyn identify bond formation and dissociation events primarily from geometric or connectivity criteria~\cite{Dontgen_2015ChemTraYzer,Zeng_2020ReacNetGenerator,Zhu_2025ReaxANA,Maddipati_2026ChemXDyn}. E3D and E3D-IQA offer a complementary possibility: bond rearrangements could be monitored not only by distances or bond orders, but also by learned energetic contributions associated with atoms and atom pairs.

\section{\label{sec:methods} Methods}
\subsection{\label{ssec:IQAref}IQA Reference Data and Energy Decomposition}
We used organic molecular structures composed of 4-23 atoms containing H, C, N, and O from the Transition1x dataset~\cite{Schreiner_2022}. Transition1x contains 10,072 reaction pathways, each consisting of a reactant structure, a transition-state structure, and a product structure. In this work, IQA labels were generated for 4-9 atom structures and for a smaller 20-23 atom external validation set. Structures containing 10-19 atoms were used only as energy/force-labeled data in mixed training experiments.

For the 4-9 atom IQA-labeled subset, the data were split by reaction pathway rather than by individual structure. Therefore, if the reactant structure of a given pathway was assigned to the training set, the corresponding transition-state and product structures were also assigned to the same split. This avoids leakage between reaction states. The 4-9 atom subset contained 1,287 training structures, 159 validation structures, and 165 test structures, corresponding to 429, 53, and 55 reaction pathways, respectively. The 10-19 atom E/F-only subset contained 22,653 training structures, 2,817 validation structures, and 2,865 test structures. The 20-23 atom IQA-labeled external validation set contained 91 reactant, 91 transition-state, and 91 product structures.

Reference electronic-structure calculations for the IQA-labeled data were performed with Psi4 at the B3LYP/def2-TZVPPD level~\cite{Becke_1993,Lee_1988,Weigend_2005,Rappoport_2010}, followed by AIMAll analysis to obtain QTAIM atomic basins and IQA energy terms~\cite{Smith_2020,Keith_2019,Maxwell_2016}. Although our previous E3D work used a different density-functional level, B3LYP was chosen here because of compatibility with the IQA workflow used in AIMAll.

For each structure, IQA decomposes the total energy into an intra-atomic term $E_{\mathrm{intra}}(i)$ and an interatomic term $E_{\mathrm{inter}}(i,j)$, as shown in eq 2. The IQA framework can in principle further decompose these terms into kinetic, electron-nuclear, electron-electron, exchange, and correlation contributions. The present study focuses on the total intra-atomic and interatomic IQA terms, while leaving this finer decomposition as a future extension.

The total energy used for model training and evaluation was the cohesive energy obtained by subtracting isolated-atom energies from the electronic energy. The IQA intra-atomic energy was treated consistently by subtracting the isolated-atom intra-atomic reference:
\begin{eqnarray}
\Delta E_{\mathrm{intra}}(i)
  = E_{\mathrm{intra}}(i) - E_{\mathrm{atom}}^{\mathrm{iso}}(Z_i),
\label{eq:intraiso}
\end{eqnarray}
where $Z_i$ denotes the chemical species of atom $i$, and $E_{\mathrm{atom}}^{\mathrm{iso}}(Z_i)$ is the isolated-atom intra-atomic energy for that species. In the remainder of the paper, $E_{\mathrm{intra}}$ refers to this isolated-atom-referenced quantity. With this convention, $E_{\mathrm{intra}}$ is typically positive and can be interpreted as an atomic deformation or strain-like contribution.

\subsection{E3D-IQA Model}
\label{ssec:E3D-IQA}
E3D-IQA was implemented on top of Allegro~\cite{Musaelian_2023}. Allegro constructs equivariant edge features for atom pairs within a local cutoff and maps these edge features to scalar edge-energy contributions. In this work, the cutoff radius was set to 5.0 Å for all main E3D-IQA models.

Let $\mathbf{x}_{ij}$ denote the learned edge feature for atom pair $(i,j)$. The original edge-energy route is retained and written as
\begin{eqnarray}
D_{ij}^{\mathrm{ML}} = \mathrm{MLP}_{\mathrm{edge}}(\mathbf{x}_{ij}).
\end{eqnarray}
This quantity is treated as the model's latent pair-energy contribution. E3D-IQA adds a node-level route by aggregating the edge features around each atom,
\begin{eqnarray}
\mathbf{x}_{i} = \sum_{j \in \mathcal{N}(i; r_{\mathrm{cut}})} \mathbf{x}_{ij},
\end{eqnarray}
and mapping the resulting node feature to an intra-atomic energy,
\begin{eqnarray}
E_{\mathrm{intra}}^{\mathrm{ML}}(i)
  = \mathrm{MLP}_{\mathrm{node}}(\mathbf{x}_{i}).
\end{eqnarray}
The total model energy is then
\begin{eqnarray}
E^{\mathrm{ML}}
  = \sum_i E_{\mathrm{intra}}^{\mathrm{ML}}(i)
  + \sum_i \sum_{j>i} D_{ij}^{\mathrm{ML}}.
\end{eqnarray}
Both the node and edge routes therefore use the same local neighbor graph and the same cutoff. Forces are obtained by differentiating the total energy with respect to atomic coordinates,
\begin{eqnarray}
\mathbf{F}_i^{\mathrm{ML}}
  = - \frac{\partial E^{\mathrm{ML}}}{\partial \mathbf{R}_i},
\end{eqnarray}
where $\mathbf{R}_i$ denotes the Cartesian coordinate vector of atom $i$. For analysis, the implementation can also output the force components arising from the node and edge energy paths. The force used for training, however, is always the total force from eq 8, and the sum of the node and edge force components is constructed to be consistent with this total force. To avoid introducing an additional species-dependent energy origin into the decomposition, the per-type energy shift was set to zero.

The detailed Allegro and E3D-IQA hyperparameters used in this work are summarized in the Supporting Information. The models used SiLU activation functions and were optimized with Adam~\cite{Elfwing_2017,Kingma_2017}.

\subsection{Training Targets and Model Variants}
\label{ssec:models}
All models were trained against total energies and forces. The explicit loss definition is given in Supporting Note 2. E3D-IQA models were additionally trained against IQA $E_{\mathrm{intra}}$ when that label was available. The IQA interatomic energy $E_{\mathrm{inter}}$ was not used as a training target in the main models. Instead, the learned latent edge energy $D_{ij}^{\mathrm{ML}}$ was compared with IQA $E_{\mathrm{inter}}$ after training.

The model variants are denoted as follows. Baseline-4-9 is the original Allegro baseline trained on 4-9 atom structures using only energy and force labels. Baseline-4-19 is the corresponding Allegro baseline trained on both the 4-9 atom and 10-19 atom energy/force data. IQA-4-9-$\lambda$ denotes an E3D-IQA model trained on the 4-9 atom IQA-labeled subset, where $\lambda$ is the loss coefficient for the $E_{\mathrm{intra}}$ term. IQA+EF-4-19-$\lambda$0.1 denotes an E3D-IQA model trained on the 4-9 atom IQA-labeled subset together with the 10-19 atom energy/force-only subset.

For the IQA-4-9-$\lambda$ series, the $E_{\mathrm{intra}}$ loss coefficient was varied systematically over $\lambda$ = 0.0, 0.01, 0.05, 0.1, 1.0, and 10.0. The $\lambda$ = 0.0 model serves as a negative control: it contains the E3D-IQA architecture, including the node-energy path, but the node energy is not supervised by IQA labels. This model tests whether an IQA-like decomposition emerges from energy and force training alone.

In the mixed IQA+EF-4-19-$\lambda$0.1 training, the 4-9 atom structures contributed total energy, force, and $E_{\mathrm{intra}}$ losses, whereas the 10-19 atom structures contributed only total energy and force losses. The $E_{\mathrm{intra}}$ loss was skipped for structures without IQA labels. This setting mimics a practical semi-supervised scenario in which a small number of expensive IQA labels are combined with a larger set of standard energy/force data.

\subsection{Evaluation Metrics}
\label{ssec:metrics}
Model accuracy was evaluated using mean absolute error (MAE). Total energy errors are reported in eV per structure, force errors in eV Å$^{-1}$, and decomposition-energy errors in eV per atom or per pair. For comparison with conventional MLIP metrics, per-atom energy MAEs are also provided in the Supporting Information. For E3D-IQA models, $E_{\mathrm{intra}}^{\mathrm{ML}}$ was compared with the IQA $E_{\mathrm{intra}}$ reference, and $D_{ij}^{\mathrm{ML}}$ was compared with the IQA $E_{\mathrm{inter}}(i,j)$ reference.

Because Allegro and E3D-IQA use a finite cutoff, $D_{ij}^{\mathrm{ML}}$ exists only for atom pairs included in the model graph. IQA $E_{\mathrm{inter}}(i,j)$, by contrast, is defined for all atom pairs. Therefore, the main $E_{\mathrm{inter}}$ parity plots and MAEs were computed for atom pairs that are present as model edges. The contribution from cutoff-excluded IQA pairs was analyzed separately.

For pair-resolved error analysis, the $E_{\mathrm{inter}}$ absolute error was separated into three categories: pairs with negative reference and negative prediction, pairs with positive reference but negative prediction, and all other cases. This decomposition helps distinguish scale errors in attractive or bonding interactions from sign-allocation errors in positive or destabilizing IQA pair terms. Parity plots were used together with MAE to assess correlations, systematic deviations from the $y=x$ line, sign mismatches, and scale errors. Force parity plots were constructed from Cartesian force components.

\section{\label{sec:results}Results and Discussion}
\subsection{\label{ssec:framework}Overview of the E3D-IQA Framework}
Figure 1 summarizes the E3D-IQA concept. The model retains the Allegro edge-energy route, which provides a latent pair-energy contribution $D_{ij}^{\mathrm{ML}}$, and adds a node-energy route trained against IQA $E_{\mathrm{intra}}$. The total energy is the sum of the node and edge contributions, and forces are obtained from this total energy. The central question is whether the IQA-labeled node supervision can guide the latent edge-energy representation toward an IQA-like interatomic energy without directly training against $E_{\mathrm{inter}}$.

This study treats E3D-IQA primarily as a diagnostic framework. The aim is not only to reproduce total energies and forces, but also to determine whether the internal energy allocation learned by the MLIP is chemically meaningful. In this sense, the model's failures are also informative: they reveal which latent components require additional labels, constraints, or architectural bias.

\subsection{\label{ssec:Eintra}Intra-Atomic Supervision Is Required for IQA-Like Decomposition}
Table 1 summarizes the MAEs of the baseline Allegro model and the IQA-4-9-$\lambda$ series for total energy, forces, $E_{\mathrm{intra}}$, and $E_{\mathrm{inter}}$. The baseline model provides the reference accuracy of an ordinary Allegro potential trained only on total energies and forces. On the validation and test sets, Baseline-4-9 achieved energy MAEs of 0.890 and 0.942 eV and force MAEs of 0.077 and 0.083 eV Å$^{-1}$, respectively.

The negative-control model IQA-4-9-$\lambda$0 achieved E/F errors similar to Baseline-4-9, with validation/test energy MAEs of 0.902/0.955 eV and force MAEs of 0.077/0.082 eV Å$^{-1}$. However, the same model gave very large decomposition errors: validation/test $E_{\mathrm{intra}}$ MAEs of 12.210/12.291 eV and $E_{\mathrm{inter}}$ MAEs of 4.566/4.592 eV. This shows that adding node and edge energy channels to the architecture is not sufficient for an IQA-like decomposition to emerge from total-energy and force training alone.

The molecular example in Figure 2(a,b) further illustrates this failure mode. In the IQA reference, the selected C, C, and O atoms have positive intra-atomic energies of 5.7, 16.9, and 5.8 eV, respectively. In contrast, IQA-4-9-$\lambda$0 assigns negative node energies of -4.9, -5.0, and -3.3 eV to the corresponding atoms. The pair contributions also fail to reproduce the IQA interatomic pattern: for example, the C-C interaction of -6.6 eV in the IQA reference becomes -0.4 eV in IQA-4-9-$\lambda$0, and the two C-N interactions of -13.8 and -19.4 eV are reduced to -0.5 and -0.6 eV. Thus, the $\lambda$0 model can learn the total energy and force while using an internal energy allocation that is chemically inconsistent with IQA.

\subsection{\label{ssec:select} 
Dependence on the Intra-Atomic Loss Weight}
Because IQA-4-9-$\lambda$0 did not recover an IQA-like energy allocation, we next examined whether direct supervision of only the intra-atomic term can guide the decomposition. Introducing a nonzero $E_{\mathrm{intra}}$ loss immediately reduced the $E_{\mathrm{intra}}$ error. For example, the validation $E_{\mathrm{intra}}$ MAE decreased from 12.210 eV at $\lambda$0 to 0.442 eV at $\lambda$0.01 and to 0.229 eV at $\lambda$0.1. The test $E_{\mathrm{intra}}$ MAE likewise decreased from 12.291 eV to 0.217 eV at $\lambda$0.1.

Increasing $\lambda$ further improved $E_{\mathrm{intra}}$ itself, but at the cost of force accuracy. IQA-4-9-$\lambda$10 achieved the lowest validation/test $E_{\mathrm{intra}}$ MAEs of 0.112/0.131 eV, but its force MAEs increased to 0.171/0.177 eV Å$^{-1}$. By contrast, IQA-4-9-$\lambda$0.1 preserved force accuracy close to the baseline while still providing accurate intra-atomic energies. Its validation/test force MAEs were 0.079/0.084 eV Å$^{-1}$, close to Baseline-4-9 values of 0.077/0.083 eV Å$^{-1}$, while its $E_{\mathrm{intra}}$ errors were much smaller than those of $\lambda$0. Therefore, IQA-4-9-$\lambda$0.1 was selected as the representative balanced model.

Figure 2(c) shows that IQA-4-9-$\lambda$0.1 recovers the molecular energy allocation far more closely than IQA-4-9-$\lambda$0. For example, the reference intra-atomic values of approximately 5.7, 16.9, and 5.8 eV for the selected C, C, and O atoms are predicted as 5.6, 16.3, and 5.9 eV. Several pair terms also become close to the IQA reference: the central C-C term remains near -6.6 eV, and the two C-N terms change from reference values of -13.8 and -19.4 eV to predicted values of -12.6 and -17.2 eV. These improvements are obtained even though $E_{\mathrm{inter}}$ was not included in the loss function.

\subsection{\label{ssec:parity} 
Parity Analysis of Total and Decomposed Energies}
Figure 3 compares parity plots for Baseline-4-9 and the representative E3D-IQA model, IQA-4-9-$\lambda$0.1. For total energy and forces, IQA-4-9-$\lambda$0.1 retains accuracy comparable to Baseline-4-9. The correlation coefficients for Baseline-4-9 were 0.962 for energy and 0.831 for force, whereas those for IQA-4-9-$\lambda$0.1 were 0.959 and 0.831, respectively. Thus, adding IQA intra-atomic supervision did not substantially degrade the ordinary E/F prediction accuracy in the 4-9 atom validation and test data.

The decomposition parity plots provide the key diagnostic information. $E_{\mathrm{intra}}$ is directly supervised and shows an almost one-to-one relationship with the IQA reference, with a correlation coefficient of 0.999. More importantly, $D_{ij}^{\mathrm{ML}}$ also shows a meaningful correlation with IQA $E_{\mathrm{inter}}$, with a correlation coefficient of 0.933, despite the absence of direct pair-energy supervision. This supports the central premise of E3D-IQA: constraining the one-body side of the energy decomposition can guide the remaining latent edge energy toward an IQA-like pair representation.

At the same time, Figure 3 reveals systematic deviations in the pair-energy allocation. Some weakly negative IQA pair terms are mapped to $D_{ij}^{\mathrm{ML}}$ values close to zero, and some positive or destabilizing IQA pair terms are not reproduced with the correct sign. To verify that this behavior is not dominated by missing long-range pairs, we estimated the cutoff-excluded contribution by assigning $D_{ij}^{\mathrm{ML}}=0$ outside the model graph. For IQA-4-9-$\lambda$0.1 on the 4-9 atom validation and test data, the edge-pair $E_{\mathrm{inter}}$ MAE was 2.098 eV, whereas the outside-cutoff zero-assignment MAE was 0.491 eV. The main $E_{\mathrm{inter}}/D_{ij}^{\mathrm{ML}}$ discrepancy in Figure 3(b) therefore originates from latent energy allocation within the modeled graph, rather than from omitted cutoff-excluded IQA interactions. Increasing the cutoff to 5.5 Å did not monotonically improve the correspondence (Figure S1).

\subsection{\label{ssec:states} 
Reaction-State Dependence of Energy-Allocation Errors}
We next examined whether the decomposition errors depend on reaction state. Figure 4 summarizes MAEs for the 4-9 atom validation and test structures separated into Reaction, Transition, and Product states. The total-energy MAEs were 0.949, 1.233, and 0.711 eV for Reaction, Transition, and Product, respectively. Force errors showed a stronger state dependence, with MAEs of 0.030, 0.123, and 0.091 eV Å$^{-1}$, indicating that transition-state structures are the most challenging for force prediction.

The intra-atomic and interatomic terms behave differently across reaction states. $E_{\mathrm{intra}}$ MAEs were 0.104, 0.311, and 0.253 eV for Reaction, Transition, and Product, respectively. Thus, the intra-atomic term is more sensitive to reaction-state changes, likely reflecting atomic deformation and electronic rearrangement during bond formation, bond breaking, and product relaxation. By contrast, $E_{\mathrm{inter}}$ MAEs remained nearly constant across the three states: 2.111, 2.116, and 2.065 eV.

The decomposed $E_{\mathrm{inter}}$ error further clarifies the source of the pair-energy discrepancy. For Reaction, Transition, and Product states, the absolute-error contributions from negative reference and negative prediction were 1.244, 1.228, and 1.193 eV, respectively. The contributions from positive reference but negative prediction were also comparable, 0.829, 0.855, and 0.827 eV. The remaining "other" category was small, 0.038, 0.033, and 0.045 eV. Cases where both the reference and prediction are positive are included in this small "other" contribution, indicating that the present E3D-IQA model rarely expresses positive IQA pair terms explicitly. These results suggest that $E_{\mathrm{inter}}$ errors arise from both sign-allocation errors for destabilizing pair terms and scale errors for attractive pair terms. Because the total energy is a sum of node and edge contributions, reducing the sign-allocation error may also change the scale error through compensation between $E_{\mathrm{intra}}$ and $D_{ij}^{\mathrm{ML}}$.

A molecular example along a Reaction, Transition, and Product sequence is provided in Figure S2. The example illustrates that the learned intra-atomic energies remain close to the IQA reference across the reaction sequence. The Reaction structure retains the main C-H, C-C, C-N, and N-O pair-energy pattern, whereas the Transition and Product structures show larger magnitude errors for N/O-centered rearranging pairs, including N-N, N-O, and C-N terms. Thus, the pair-energy error is not uniformly distributed over the molecule, but is localized to specific pairwise allocations. The pair-type dependence of these errors is analyzed next.

\subsection{\label{ssec:species} 
Species- and Pair-Resolved Error Analysis}
Figure 5 further decomposes the decomposition-energy errors: $E_{\mathrm{intra}}$ errors are grouped by atomic species, whereas $E_{\mathrm{inter}}/D_{ij}^{\mathrm{ML}}$ errors are grouped by pair type. The species-resolved $E_{\mathrm{intra}}$ errors were small for all elements, with MAEs of 0.107 eV for H, 0.216 eV for O, 0.288 eV for N, and 0.312 eV for C. This indicates that the effect of intra-atomic supervision is retained across all H, C, N, and O species in the organic subset.

The pair-resolved $E_{\mathrm{inter}}$ errors were substantially larger and more heterogeneous. O-O had the largest MAE of 5.697 eV, almost entirely from positive reference terms predicted as negative $D_{ij}^{\mathrm{ML}}$ values (5.683 eV). N-O showed a similar pattern, with a total MAE of 4.213 eV and a positive-reference-to-negative-prediction contribution of 3.851 eV. These pair types are not necessarily covalent bonds; all atom pairs within the model cutoff are included. Therefore, large O-O and N-O errors likely reflect close nonbonded or destabilizing contacts that require positive IQA pair terms.

Other pair types show a different balance of errors. C-O and C-N had MAEs of 3.937 and 3.540 eV, dominated by cases where both reference and prediction were negative, with contributions of 3.838 and 3.350 eV, respectively. In these cases, the sign is often correct but the magnitude is not. C-C pairs had a mixed character, with a total MAE of 3.973 eV, including 1.417 eV from negative-reference/negative-prediction cases and 2.548 eV from positive-reference/negative-prediction cases. These results show that a single pair type can contain chemically different situations, including covalent bonds, stretched bonds, and nonbonded contacts within the cutoff. The remaining pair types, such as N-N, C-H, H-O, H-N, and H-H, followed the same qualitative error modes, namely sign-allocation errors for positive IQA pair terms and magnitude errors for negative pair terms.

This analysis highlights the value of IQA-based diagnostics. From total energy and force errors alone, it would be difficult to identify whether the model fails because of attractive pair scaling, missing positive pair contributions, or compensating one-body/two-body allocation. The IQA decomposition exposes these error modes directly and suggests concrete improvements, such as weak pair-energy supervision, sign-aware constraints, or distance-dependent inductive bias.

Further decomposition of the parity plots in Figure 3(b) by element and atom-pair type provides a practical tool for tracing aggregate decomposition errors to specific chemical classes, reference-energy ranges, and sign-mismatch patterns (Figures S3 and S4).

\subsection{\label{ssec:transfer} 
Transfer with Energy/Force-Only Data}
Many large atomistic datasets do not contain IQA decomposition labels, but they do contain standard energy and force labels. We therefore tested whether E/F-only structures can improve E3D-IQA transfer while retaining the IQA-derived interpretability learned from the smaller labeled subset. In IQA+EF-4-19-$\lambda$0.1, the 4-9 atom data provide E/F/$E_{\mathrm{intra}}$ labels, whereas the 10-19 atom data provide only E/F labels. The resulting performance is summarized in Table 2.

For compact comparison, the 4-9 atom results in Table 2 combine the validation and test subsets. Adding E/F-only data improved both ordinary prediction accuracy and decomposition diagnostics. On the 4-9 atom data, IQA+EF-4-19-$\lambda$0.1 reduced the total-energy MAE from 0.964 to 0.422 eV and the force MAE from 0.082 to 0.063 eV Å$^{-1}$ relative to IQA-4-9-$\lambda$0.1. On the 20-23 atom data, the improvement was larger: the total-energy MAE decreased from 1.656 to 0.706 eV, and the force MAE decreased from 0.249 to 0.036 eV Å$^{-1}$. IQA+EF-4-19-$\lambda$0.1 also achieved E/F accuracy comparable to Baseline-4-19 on both datasets, while retaining IQA-type decomposition diagnostics.

The mixed model also improved the IQA-related diagnostics. On the 4-9 atom data, the $E_{\mathrm{intra}}$ MAE decreased from 0.223 to 0.143 eV, and the $E_{\mathrm{inter}}$ MAE decreased from 2.098 to 1.890 eV. On the 20-23 atom data, $E_{\mathrm{intra}}$ improved from 0.291 to 0.268 eV, and $E_{\mathrm{inter}}$ improved from 0.326 to 0.247 eV. Thus, E/F-only data can improve not only total energy and force transfer, but also the latent decomposition learned from a smaller IQA-labeled subset.

The molecular example in Figure 2(d) is consistent with this trend. IQA+EF-4-19-$\lambda$0.1 preserves the IQA-like intra-atomic energy allocation and maintains pair energies close to the reference for several chemically important pairs. For example, the selected C, C, and O intra-atomic energies are predicted as 5.6, 16.9, and 5.8 eV, close to the IQA values of 5.7, 16.9, and 5.8 eV.

\subsection{\label{ssec:limitations} 
Current Limitations}
E3D-IQA should not be viewed as a complete surrogate model for IQA at this stage. Rather, it provides a way to examine whether the internal energy allocation of an MLIP is consistent with a quantum-chemical decomposition, and to identify which model components require additional information. The present results show that total-energy and force accuracy alone does not guarantee a meaningful one-body/two-body decomposition.

The first limitation is that $E_{\mathrm{inter}}$ was not directly supervised. This staged design was intentional: because the total energy is already the sum of one-body and two-body contributions, simultaneously constraining total energy, forces, $E_{\mathrm{intra}}$, and $E_{\mathrm{inter}}$ could overconstrain the model or introduce training instability. The E3D result also motivated the possibility that pair-like information could emerge in the latent edge energy. The present study shows that this emergence is partial but incomplete.

The second limitation concerns the signs and scales of the decomposed terms. Because the isolated-atom-referenced $E_{\mathrm{intra}}$ is typically positive, the node-energy path may require explicit positivity or scale-aware inductive bias for more stable decomposition. For the pair-energy path, the most important remaining problem is the treatment of positive IQA interatomic terms and weak interactions. These errors are not evidence that MLIPs cannot learn repulsion; rather, they indicate that the internal allocation of the total energy does not automatically match the IQA allocation. This limitation points to the need for additional constraints, such as weak or partial $E_{\mathrm{inter}}$ supervision, sign-aware losses, distance-dependent constraints, or long-range terms.

Implementing direct or partial $E_{\mathrm{inter}}$ supervision is also a practical limitation. Unlike $E_{\mathrm{intra}}$, which is an atom-wise label, $E_{\mathrm{inter}}$ is a pair-wise label. Therefore, extxyz or database formats, data loaders, Allegro edge outputs, and loss functions must be extended to handle pair labels consistently.

\section{\label{sec:conclusion}Conclusions}
We introduced E3D-IQA, an Allegro-based MLIP framework that connects latent edge-energy decomposition with IQA quantum-chemical energy decomposition. The model retains the original Allegro edge-energy route and adds a node-energy route trained against IQA $E_{\mathrm{intra}}$. The IQA interatomic energy is not used as a training target; instead, the learned latent edge energy is evaluated as a post-training representation corresponding to IQA $E_{\mathrm{inter}}$.

The results show that intra-atomic supervision is essential. A model with the E3D-IQA architecture but without $E_{\mathrm{intra}}$ supervision can reproduce total energies and forces, yet fails to produce an IQA-like decomposition. Once $E_{\mathrm{intra}}$ supervision is introduced, the node energy closely follows the IQA intra-atomic reference, and the latent edge energy exhibits a meaningful correspondence with IQA $E_{\mathrm{inter}}$ even without direct pair-energy supervision.

The IQA decomposition also exposes internal model errors that are hidden in ordinary energy and force metrics. In particular, positive or destabilizing IQA pair terms are often assigned negative latent edge energies, and attractive pair terms can have incorrect magnitudes. These findings turn the apparent weaknesses of the model into actionable diagnostics: they indicate where additional labels, sign-aware constraints, long-range terms, or pair-wise data formats are required.

Adding 10-19 atom structures with energy and force labels but without IQA labels improved transfer to larger molecules and reduced both ordinary E/F errors and IQA-related decomposition errors. This result suggests that expensive IQA labels do not need to be available for all structures. Instead, a smaller IQA-labeled subset can be combined with larger standard E/F datasets to guide and regularize latent energy decomposition.

Overall, E3D-IQA provides a route for diagnosing and guiding latent energy allocation in MLIPs using quantum-chemical decomposition labels. Although the present study focused on H/C/N/O organic reactions, the underlying IQA framework is more general and is not restricted to covalent molecules. Extending this strategy to ionic systems, metallic systems, molecule-surface interfaces, and broader reactive datasets, together with pair-wise IQA supervision and sign-aware constraints, may enable energy-resolved MLIPs that support both accurate simulation and chemically interpretable analysis of reactive processes.

\section*{Author Contributions}
{\bf{Kohei Shimamura:}} Conceptualization; methodology; software; formal analysis; data curation; visualization; writing--original draft; writing--review and editing.
{\bf{Ken-ichi Nomura:}} Methodology; software; writing--review and editing.

\section*{Acknowledgments}
This work was supported by Taiwan Semiconductor Manufacturing Company (TSMC) / Japan Advanced Semiconductor Manufacturing (JASM).

\section*{Conflict of Interest}
The authors declare that they have no known competing financial interests or personal relationships that could have appeared to influence the work reported in this paper. 
However, this study was funded by Taiwan Semiconductor Manufacturing Company (TSMC) and/or its manufacturing subsidiary, Japan Advanced Semiconductor Manufacturing (JASM).

\section*{Data Availability Statement}
The code and data supporting the findings of this study will be made publicly available in a GitHub repository. 

\clearpage
\bibliographystyle{unsrtnat}
\bibliography{Manuscript}

\clearpage
\begin{table}[H]
\centering
\caption{MAEs of Baseline-4--9 and the IQA-4--9-$\lambda$ series. Energy and decomposition-energy errors are in eV, and force errors are in eV \AA$^{-1}$. All models were trained against total energies $E$ and forces $F$. For E3D-IQA models with $\lambda_{\mathrm{intra}}>0$, IQA $E_{\mathrm{intra}}$ was included as an additional training target. IQA $E_{\mathrm{inter}}$ was not included in the loss for any model and was used only for post-training evaluation of $D_{ij}^{\mathrm{ML}}$ on model edges.}
\label{tab:lambda_series_mae}
\resizebox{\textwidth}{!}{%
\begin{tabular}{lccccccccccccc}
\toprule
 &  & \multicolumn{4}{c}{Train} & \multicolumn{4}{c}{Validation} & \multicolumn{4}{c}{Test} \\
\cmidrule(lr){3-6}\cmidrule(lr){7-10}\cmidrule(lr){11-14}
Model & $\lambda_{\mathrm{intra}}$ &
$E$ & $F$ & $E_{\mathrm{intra}}$ & $E_{\mathrm{inter}}$ &
$E$ & $F$ & $E_{\mathrm{intra}}$ & $E_{\mathrm{inter}}$ &
$E$ & $F$ & $E_{\mathrm{intra}}$ & $E_{\mathrm{inter}}$ \\
\midrule
Baseline-4--9 & -- & 0.837 & 0.057 & -- & -- & 0.890 & 0.077 & -- & -- & 0.942 & 0.083 & -- & -- \\
IQA-4--9-$\lambda 0$ & 0 & 0.847 & 0.058 & 12.619 & 4.909 & 0.902 & 0.077 & 12.210 & 4.566 & 0.955 & 0.082 & 12.291 & 4.592 \\
IQA-4--9-$\lambda 0.01$ & 0.01 & 0.871 & 0.058 & 0.431 & 2.337 & 0.920 & 0.076 & 0.442 & 2.143 & 0.980 & 0.077 & 0.457 & 2.161 \\
IQA-4--9-$\lambda 0.05$ & 0.05 & 0.815 & 0.057 & 0.230 & 2.233 & 0.871 & 0.080 & 0.261 & 2.059 & 0.934 & 0.086 & 0.245 & 2.061 \\
IQA-4--9-$\lambda 0.1$ & 0.1 & 0.895 & 0.060 & 0.201 & 2.281 & 0.945 & 0.079 & 0.229 & 2.090 & 0.983 & 0.084 & 0.217 & 2.105 \\
IQA-4--9-$\lambda 1$ & 1 & 0.758 & 0.061 & 0.090 & 2.234 & 0.813 & 0.106 & 0.137 & 2.078 & 0.870 & 0.118 & 0.150 & 2.064 \\
IQA-4--9-$\lambda 10$ & 10 & 0.697 & 0.064 & 0.031 & 2.185 & 0.801 & 0.171 & 0.112 & 2.022 & 0.817 & 0.177 & 0.131 & 2.015 \\
\bottomrule
\end{tabular}%
}
\end{table}

\begin{table}[H]
\centering
\caption{Effect of adding 10--19 atom E/F-only data. The 4--9 columns summarize the validation and test data, and the 20--23 columns summarize the external validation data. Energy and decomposition-energy errors are in eV, and force errors are in eV \AA$^{-1}$.}
\label{tab:ef_only_transfer_mae}
\resizebox{\textwidth}{!}{%
\begin{tabular}{lccccccccc}
\toprule
 &  & \multicolumn{4}{c}{4--9 atom data} & \multicolumn{4}{c}{20--23 atom data} \\
\cmidrule(lr){3-6}\cmidrule(lr){7-10}
Model & $\lambda_{\mathrm{intra}}$ &
$E$ & $F$ & $E_{\mathrm{intra}}$ & $E_{\mathrm{inter}}$ &
$E$ & $F$ & $E_{\mathrm{intra}}$ & $E_{\mathrm{inter}}$ \\
\midrule
Baseline-4--19 & -- & 0.287 & 0.057 & -- & -- & 0.471 & 0.034 & -- & -- \\
IQA-4--9-$\lambda 0.1$ & 0.1 & 0.964 & 0.082 & 0.223 & 2.098 & 1.656 & 0.249 & 0.291 & 0.326 \\
IQA+EF-4--19-$\lambda 0.1$ & 0.1 & 0.422 & 0.063 & 0.143 & 1.890 & 0.706 & 0.036 & 0.268 & 0.247 \\
\bottomrule
\end{tabular}%
}
\end{table}

\clearpage
\begin{figure}[H]
\begin{center}  
  \includegraphics[width=8cm]{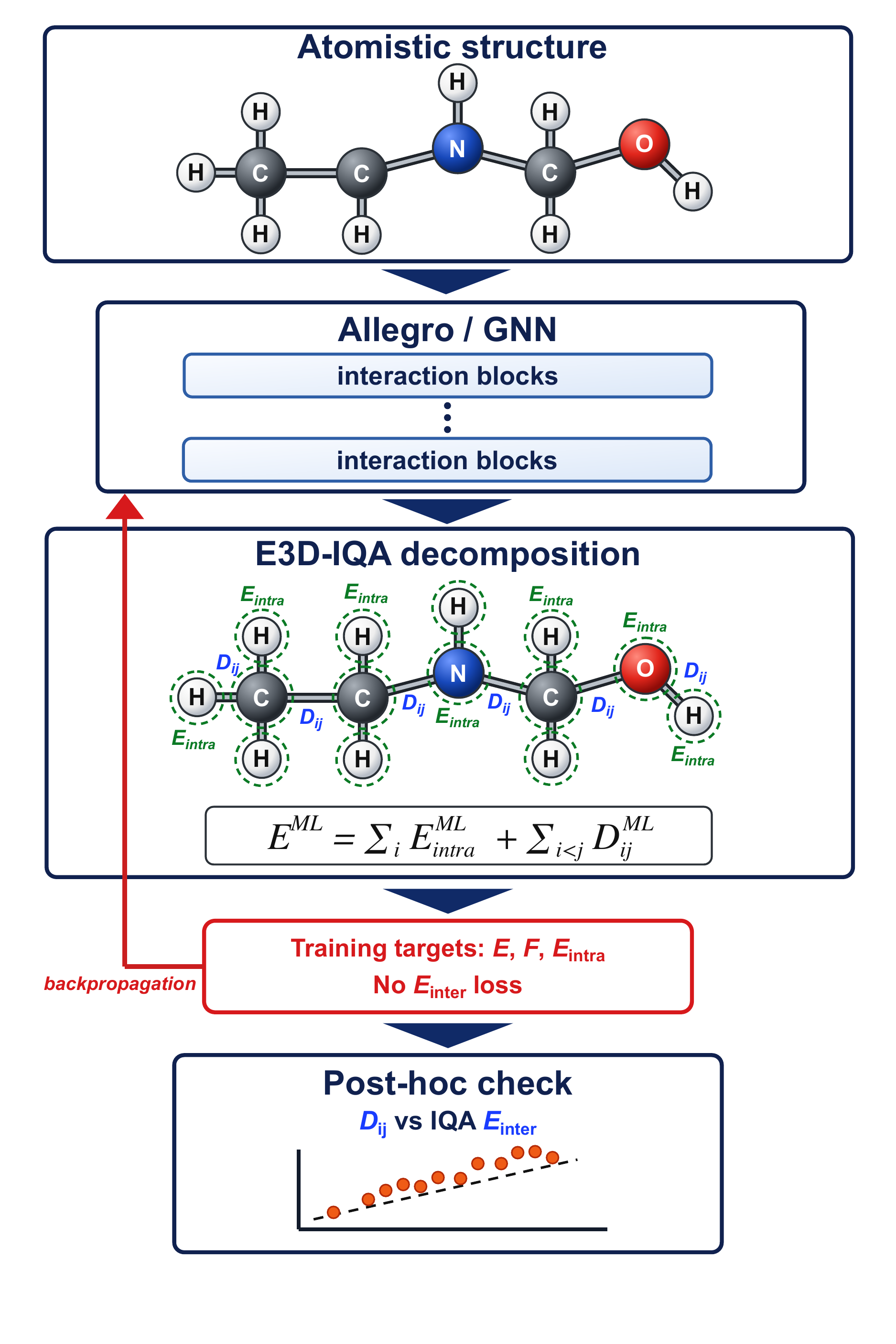}
  \caption{\label{Fig_framework}Conceptual overview of E3D-IQA. Allegro edge features are used in two routes: an edge route that produces the latent pair contribution $D_{ij}^{\mathrm{ML}}$, and a node route that aggregates edge features around each atom and predicts $E_{\mathrm{intra}}^{\mathrm{ML}}$. The model is trained against total energy, forces, and IQA $E_{\mathrm{intra}}$, whereas IQA $E_{\mathrm{inter}}$ is used only for post-training diagnosis of the latent edge energy.}
\end{center}
\end{figure}

\begin{figure}[H]
\begin{center}
  \includegraphics[width=0.95\textwidth]{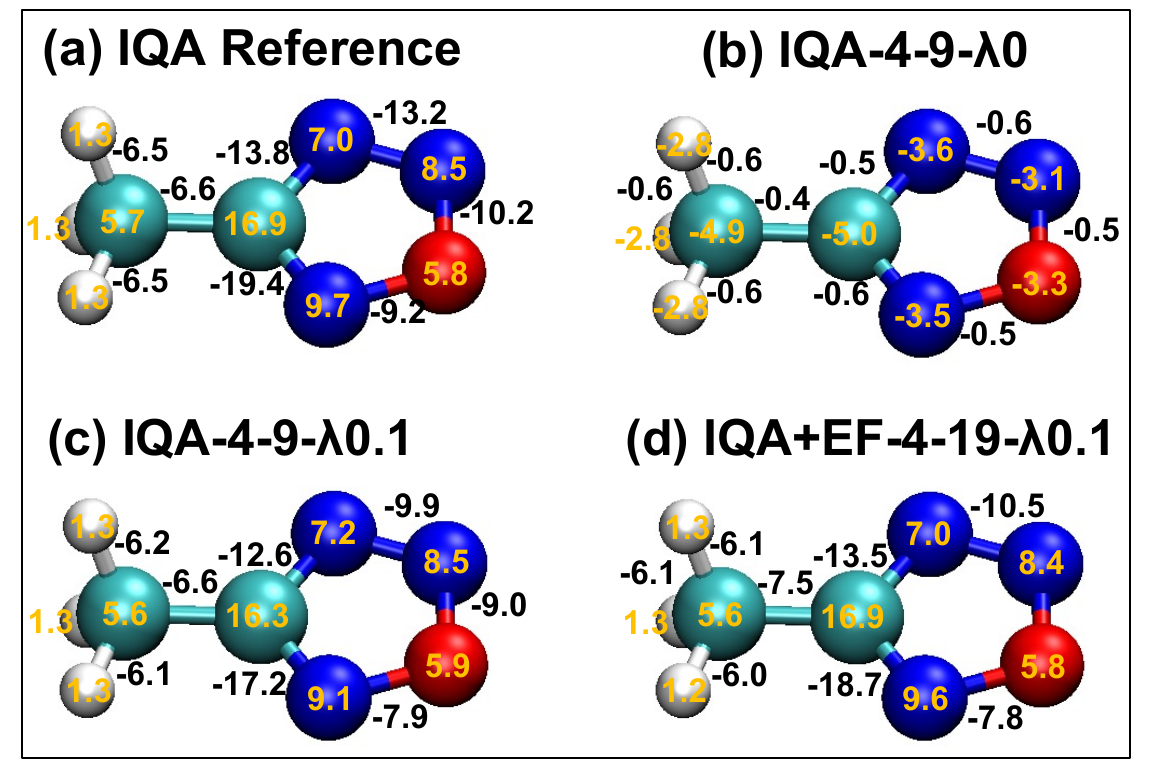}
  \caption{\label{Fig_HCNO}
  Molecular-level comparison of IQA reference decomposition and model-predicted decomposition for a representative structure. White, cyan, blue, and red spheres represent H, C, N, and O atoms, respectively. Yellow labels denote $E_{\mathrm{intra}}$, and black labels denote IQA $E_{\mathrm{inter}}$ or model-predicted latent pair energy $D_{ij}^{\mathrm{ML}}$. The unit is eV. (a) IQA reference. (b) IQA-4-9-$\lambda$0. (c) IQA-4-9-$\lambda$0.1. (d) IQA+EF-4-19-$\lambda$0.1.}
\end{center}
\end{figure}

\begin{figure}[H]
\begin{center}
  \includegraphics[width=0.95\textwidth]{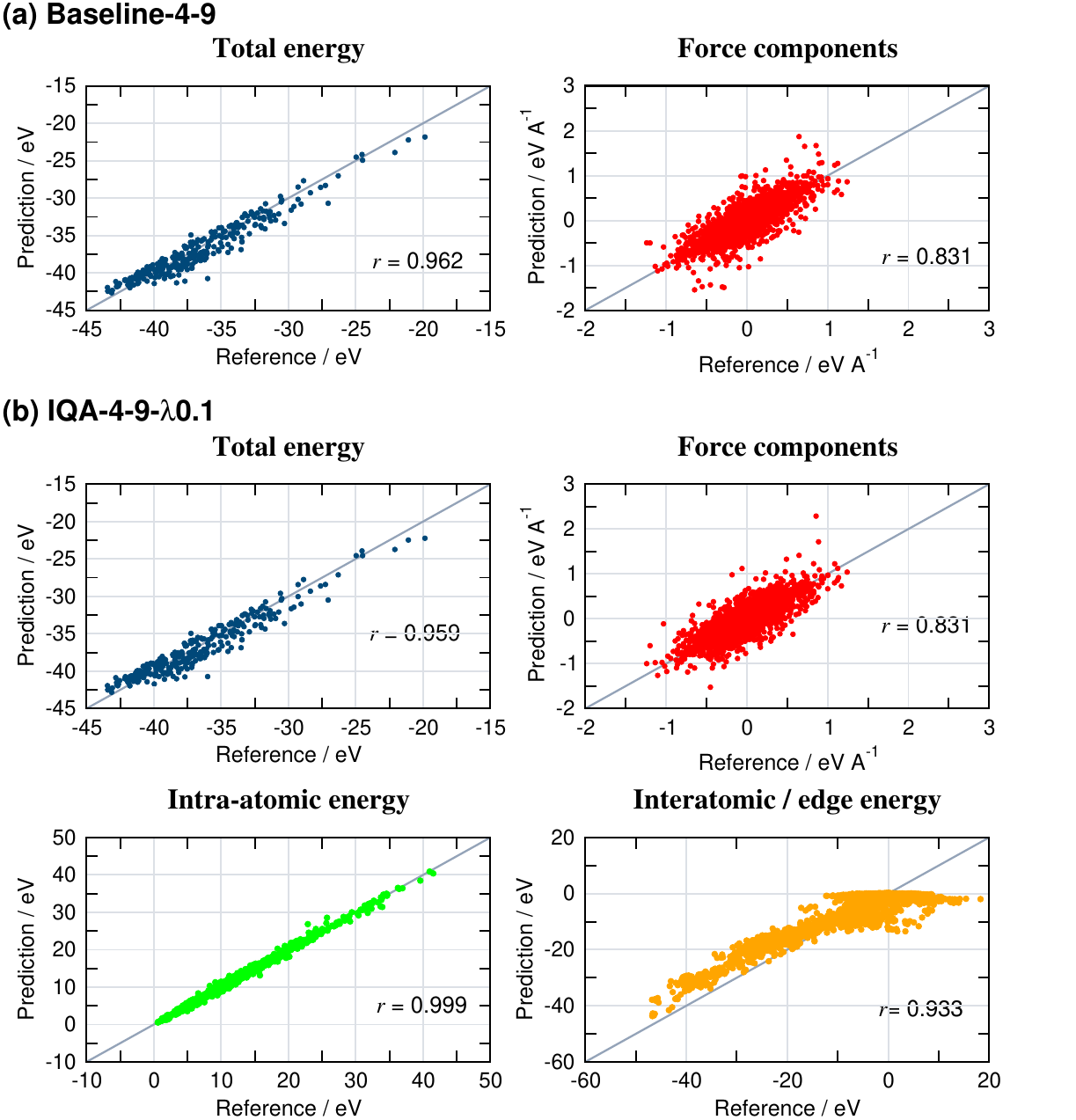}
  \caption{\label{Fig_parity}
  Parity plots for (a) Baseline-4-9 and (b) IQA-4-9-$\lambda$0.1 on the 4-9 atom validation and test data. Baseline-4-9 is evaluated for total energy and force only. IQA-4-9-$\lambda$0.1 is evaluated for total energy, force, $E_{\mathrm{intra}}$, and $E_{\mathrm{inter}}$, where the latter compares $D_{ij}^{\mathrm{ML}}$ with IQA $E_{\mathrm{inter}}$.}
\end{center}
\end{figure}

\begin{figure}[H]
\begin{center}
  \includegraphics[width=0.95\textwidth]{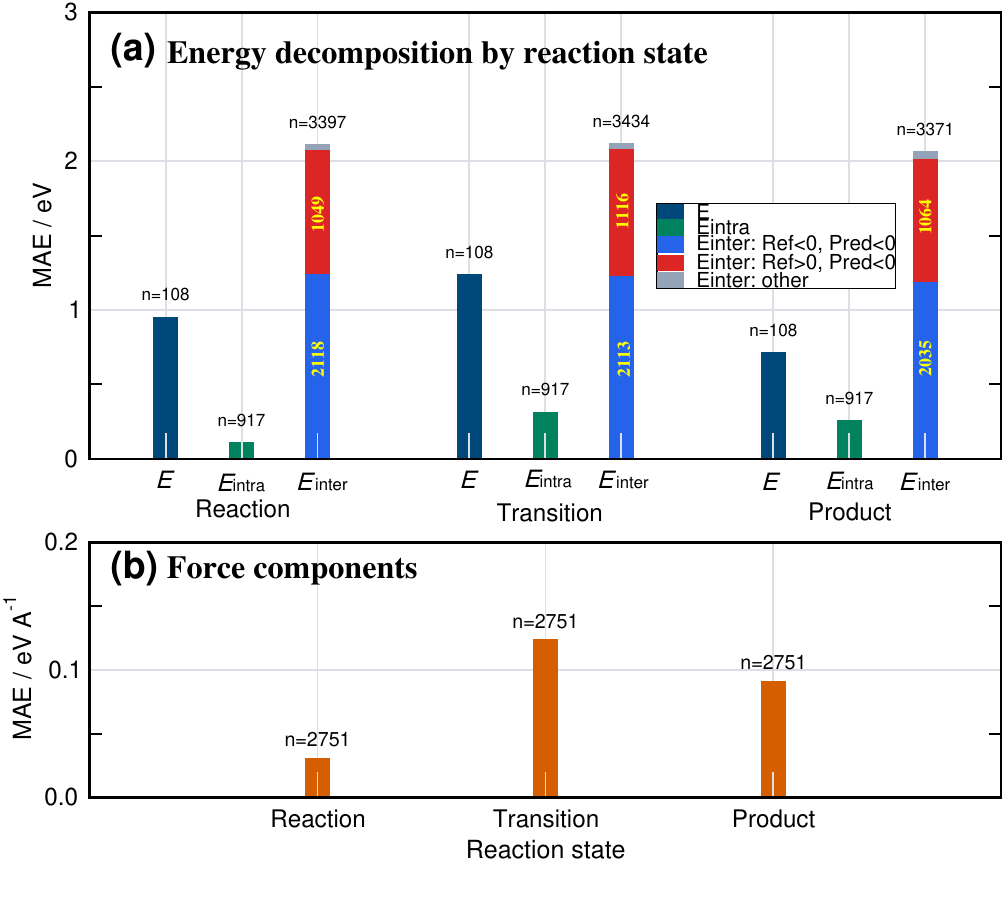}
  \caption{\label{Fig_RTP}
  Reaction-state-resolved MAEs for IQA-4-9-$\lambda$0.1 on the 4-9 atom validation and test data. The bars are ordered from left to right as Reaction, Transition, and Product states. (a) Energy-related errors. (b) Force-component errors. For $E_{\mathrm{inter}}$, the absolute-error contribution is decomposed into negative reference and negative prediction, positive reference but negative prediction, and all other cases. The value $n$ above each bar denotes the number of evaluation points used for that MAE. For the decomposed $E_{\mathrm{inter}}$ bars, the yellow numbers within the blue and red segments denote the numbers of pairwise evaluation points in the $E_{\mathrm{inter}}^{\mathrm{ref}}<0$, $D_{ij}^{\mathrm{ML}}<0$ and $E_{\mathrm{inter}}^{\mathrm{ref}}>0$, $D_{ij}^{\mathrm{ML}}<0$ categories, respectively.}
\end{center}
\end{figure}

\begin{figure}[H]
\begin{center}
  \includegraphics[width=0.95\textwidth]{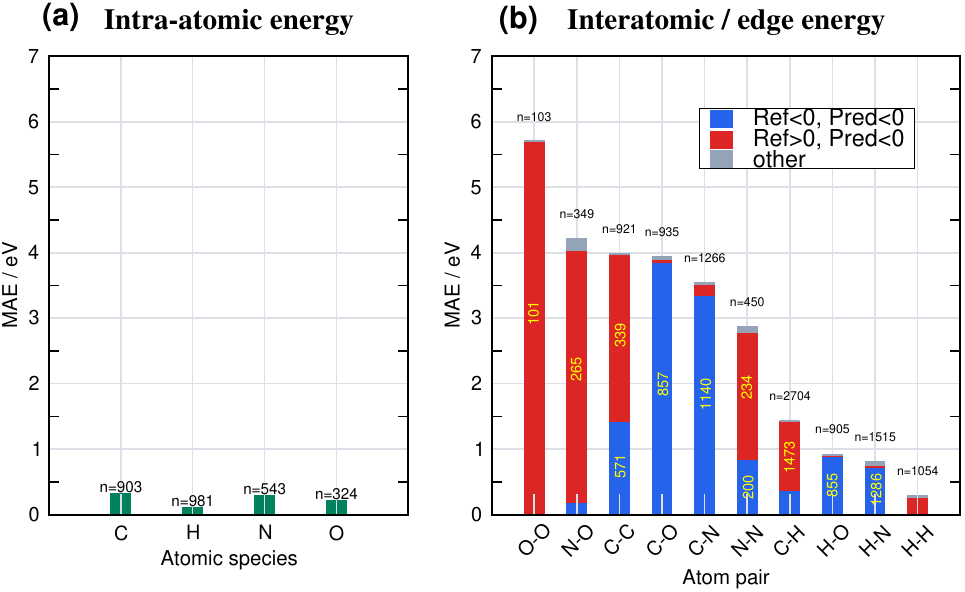}
  \caption{\label{Fig_species}
  Species- and pair-resolved MAEs for IQA-4-9-$\lambda$0.1 on the 4-9 atom validation and test data. (a) $E_{\mathrm{intra}}$ errors grouped by element. (b) $E_{\mathrm{inter}}/D_{ij}^{\mathrm{ML}}$ errors grouped by pair type and decomposed into negative reference and negative prediction, positive reference but negative prediction, and all other cases. The value $n$ above each $E_{\mathrm{intra}}$ bar denotes the number of atomic evaluation points for that element, whereas $n$ above each $E_{\mathrm{inter}}$ bar denotes the number of pairwise evaluation points for that pair type. For the decomposed $E_{\mathrm{inter}}$ bars, the yellow numbers within the blue and red segments denote the numbers of pairwise evaluation points in the $E_{\mathrm{inter}}^{\mathrm{ref}}<0$, $D_{ij}^{\mathrm{ML}}<0$ and $E_{\mathrm{inter}}^{\mathrm{ref}}>0$, $D_{ij}^{\mathrm{ML}}<0$ categories, respectively.}
\end{center}
\end{figure}

\clearpage
\input{supporting_information}

\end{document}

%% file: supporting_information.tex
\clearpage
\phantomsection
\addcontentsline{toc}{section}{Supporting Information}
\begin{center}
{\Large\bfseries Supporting Information for\\[0.4em]
``Diagnosing Latent Energy Decomposition in Machine-Learning Interatomic Potentials via Interacting Quantum Atoms''\par}
\vspace{1em}
{\large Kohei Shimamura and Ken-ichi Nomura\par}
\end{center}

\vspace{1em}
\noindent This section provides per-atom energy errors, model and training hyperparameters, the loss-function definition, cutoff-radius analysis, reaction-state-resolved molecular examples, and element- and pair-resolved parity diagnostics.

\setcounter{section}{0}
\setcounter{figure}{0}
\setcounter{table}{0}
\setcounter{equation}{0}
\renewcommand{\thesection}{S\arabic{section}}
\renewcommand{\thefigure}{S\arabic{figure}}
\renewcommand{\thetable}{S\arabic{table}}
\renewcommand{\theequation}{S\arabic{equation}}
\renewcommand{\theHsection}{SI.section.\arabic{section}}
\renewcommand{\theHfigure}{SI.figure.\arabic{figure}}
\renewcommand{\theHtable}{SI.table.\arabic{table}}
\renewcommand{\theHequation}{SI.equation.\arabic{equation}}

\clearpage
\section[Per-Atom Energy MAEs for the IQA-4-9 Lambda Series]{Per-Atom Energy MAEs for the IQA-4-9-$\lambda$ Series}
The main text reports total-energy MAEs in eV per structure to keep the total-energy error directly comparable with the one-body and pairwise energy-allocation errors. For comparison with conventional MLIP reporting, Table S1 lists the corresponding per-atom energy MAEs. The values were computed as the mean of $|E_{\mathrm{pred}}-E_{\mathrm{ref}}|/N_{\mathrm{atom}}$ over structures in each split.
\begin{table}[h]
\centering
\caption{Per-atom energy MAEs for the IQA-4--9-$\lambda$ series. The values were computed as the mean of $|E_{\mathrm{pred}}-E_{\mathrm{ref}}|/N_{\mathrm{atom}}$ over structures in each split.}
\label{tab:s_per_atom_energy_lambda_series}
\begin{tabular}{lcccc}
\toprule
Model & $\lambda_{\mathrm{intra}}$ & Train $E$ & Validation $E$ & Test $E$ \\
\midrule
Baseline-4--9 & -- & 0.101 & 0.106 & 0.112 \\
IQA-4--9-$\lambda 0$ & 0 & 0.102 & 0.107 & 0.113 \\
IQA-4--9-$\lambda 0.01$ & 0.01 & 0.105 & 0.109 & 0.116 \\
IQA-4--9-$\lambda 0.05$ & 0.05 & 0.098 & 0.103 & 0.111 \\
IQA-4--9-$\lambda 0.1$ & 0.1 & 0.108 & 0.112 & 0.117 \\
IQA-4--9-$\lambda 1$ & 1 & 0.091 & 0.096 & 0.103 \\
IQA-4--9-$\lambda 10$ & 10 & 0.084 & 0.095 & 0.097 \\
\bottomrule
\end{tabular}
\end{table}

\clearpage
\section{Per-Atom Energy MAEs for the Mixed E/F-Only Training Comparison}
Table S2 lists the per-atom energy MAEs corresponding to the model comparison in Table 2 of the main text. The 4-9 atom values combine the validation and test splits, and the 20-23 atom values correspond to the external validation set.
\begin{table}[h]
\centering
\caption{Per-atom energy MAEs for the mixed E/F-only training comparison. The 4--9 atom values combine the validation and test splits, and the 20--23 atom values correspond to the external validation set.}
\label{tab:s_per_atom_energy_mixed_training}
\begin{tabular}{lccc}
\toprule
Model & $\lambda_{\mathrm{intra}}$ & 4--9 $E$ & 20--23 $E$ \\
\midrule
Baseline-4--19 & -- & 0.034 & 0.023 \\
IQA-4--9-$\lambda 0.1$ & 0.1 & 0.114 & 0.080 \\
IQA+EF-4--19-$\lambda 0.1$ & 0.1 & 0.050 & 0.034 \\
\bottomrule
\end{tabular}
\end{table}

\clearpage
\section{Supporting Note 1. Model and Training Hyperparameters}
The main Allegro and E3D-IQA hyperparameters used in this work are summarized below. Unless otherwise noted, the same architecture and training settings were used for Baseline-4-9, Baseline-4-19, the IQA-4-9-$\lambda$ series, and IQA+EF-4-19-$\lambda$0.1. The baseline models omit the E3D-IQA node-energy loss and do not output IQA-type decomposition channels.
\begin{itemize}
\item Chemical species: H, C, N, and O.
\item Data type and precision: ASE-readable extxyz input, float32 model dtype.
\item Random seed: 456 for both data handling and model initialization.
\item Cutoff radius: 5.0 Å for the main models. The cutoff test in Figure S1 used 5.5 Å.
\item Neighbor list: generated using the same cutoff radius as the model edge graph.
\item Spherical harmonics: maximum rotation order $l_{\max}=2$.
\item Number of Allegro layers: 3.
\item Scalar features: 64.
\item Tensor features: 32.
\item Radial-chemical embedding: two-body Bessel scalar embedding with 8 Bessel functions, non-trainable Bessel basis, and polynomial cutoff power 6.
\item Scalar embedding MLP: one hidden layer, width 64, SiLU nonlinearity.
\item Allegro latent MLPs: one hidden layer, width 64, SiLU nonlinearity.
\item Readout MLP: one hidden layer, width 64, SiLU nonlinearity.
\item Parity: enabled.
\item Tensor-product path-channel coupling: enabled.
\item E3D-IQA node-energy route: edge features were summed over neighbors of each atom and passed to the node-energy MLP to produce $E_{\mathrm{intra}}^{\mathrm{ML}}$.
\item Edge-energy route: the original Allegro edge readout was retained and interpreted as the latent pair energy $D_{ij}^{\mathrm{ML}}$.
\item Edge-sum normalization: \texttt{avg\_num\_neighbors} was set from the training-data estimate of the mean number of neighbors.
\item Per-type energy shifts: fixed to 0.0.
\item Per-type energy scales: set from the training-data force RMS and kept non-trainable.
\item Loss function: total energy coefficient 1.0 and force coefficient 1.0. For E3D-IQA models, the $E_{\mathrm{intra}}$ loss coefficient was $\lambda$, with $\lambda$ = 0.0, 0.01, 0.05, 0.1, 1.0, or 10.0 for the IQA-4-9-$\lambda$ series. The representative IQA+EF-4-19 model used $\lambda$ = 0.1.
\item Energy loss normalization: per-atom energy loss was enabled during training.
\item Optimizer: Adam with learning rate $1.0 \times 10^{-3}$.
\item Weight decay: Adam default, corresponding to no explicit weight decay.
\item Batch size: 10 for training, validation, and testing.
\item Training length: 2000 epochs.
\item Checkpoint selection: the best checkpoint was selected by the validation weighted-sum metric; the last checkpoint was also saved.
\item Compilation: PyTorch compile mode was enabled for the model during training.
\end{itemize}

\clearpage
\section{Supporting Note 2. Loss Function}
The E3D-IQA model parameters were optimized using a joint loss function containing total energy, atomic forces, and the IQA intra-atomic energy when available. For a structure with $N$ atoms, the loss can be written schematically as
\begin{equation}
\begin{aligned}
\mathcal{L}
&= \lambda_E
  \left|
  \frac{E_{\mathrm{ref}} - E^{\mathrm{ML}}}{N}
  \right|^2
+ \lambda_F
  \frac{1}{3N}
  \sum_{i=1}^{N}
  \sum_{\alpha=x,y,z}
  \left|
  F_{i\alpha}^{\mathrm{ref}} - F_{i\alpha}^{\mathrm{ML}}
  \right|^2 \\
&\quad + \lambda_{\mathrm{intra}}
  \frac{1}{N}
  \sum_{i=1}^{N}
  \left|
  E_{\mathrm{intra}}^{\mathrm{ref}}(i)
  - E_{\mathrm{intra}}^{\mathrm{ML}}(i)
  \right|^2 .
\end{aligned}
\end{equation}
Here, $E^{\mathrm{ML}}$ is the total predicted energy defined in eq 7 of the main text, and $F_{i\alpha}^{\mathrm{ML}}$ is obtained by automatic differentiation of $E^{\mathrm{ML}}$ with respect to atomic coordinates. The coefficients for total energy and force were fixed to $\lambda_E=1.0$ and $\lambda_F=1.0$. The intra-atomic coefficient $\lambda_{\mathrm{intra}}$ was varied as 0.0, 0.01, 0.05, 0.1, 1.0, and 10.0 in the IQA-4-9-$\lambda$ series. The representative IQA-4-9-$\lambda$0.1 and IQA+EF-4-19-$\lambda$0.1 models used $\lambda_{\mathrm{intra}}=0.1$.

For structures without IQA labels, such as the 10-19 atom E/F-only data used in the mixed training, the $E_{\mathrm{intra}}$ term was omitted from the loss. The IQA interatomic energy $E_{\mathrm{inter}}$ and the corresponding latent edge energy $D_{ij}^{\mathrm{ML}}$ do not appear explicitly in the loss function. Therefore, the comparison between $D_{ij}^{\mathrm{ML}}$ and IQA $E_{\mathrm{inter}}$ is a post-training diagnostic of the latent edge-energy representation, not a result of direct pair-energy supervision.

\clearpage
\section{Effect of increasing the cutoff radius}
\begin{figure}[h]
\begin{center}
  \includegraphics[width=14cm]{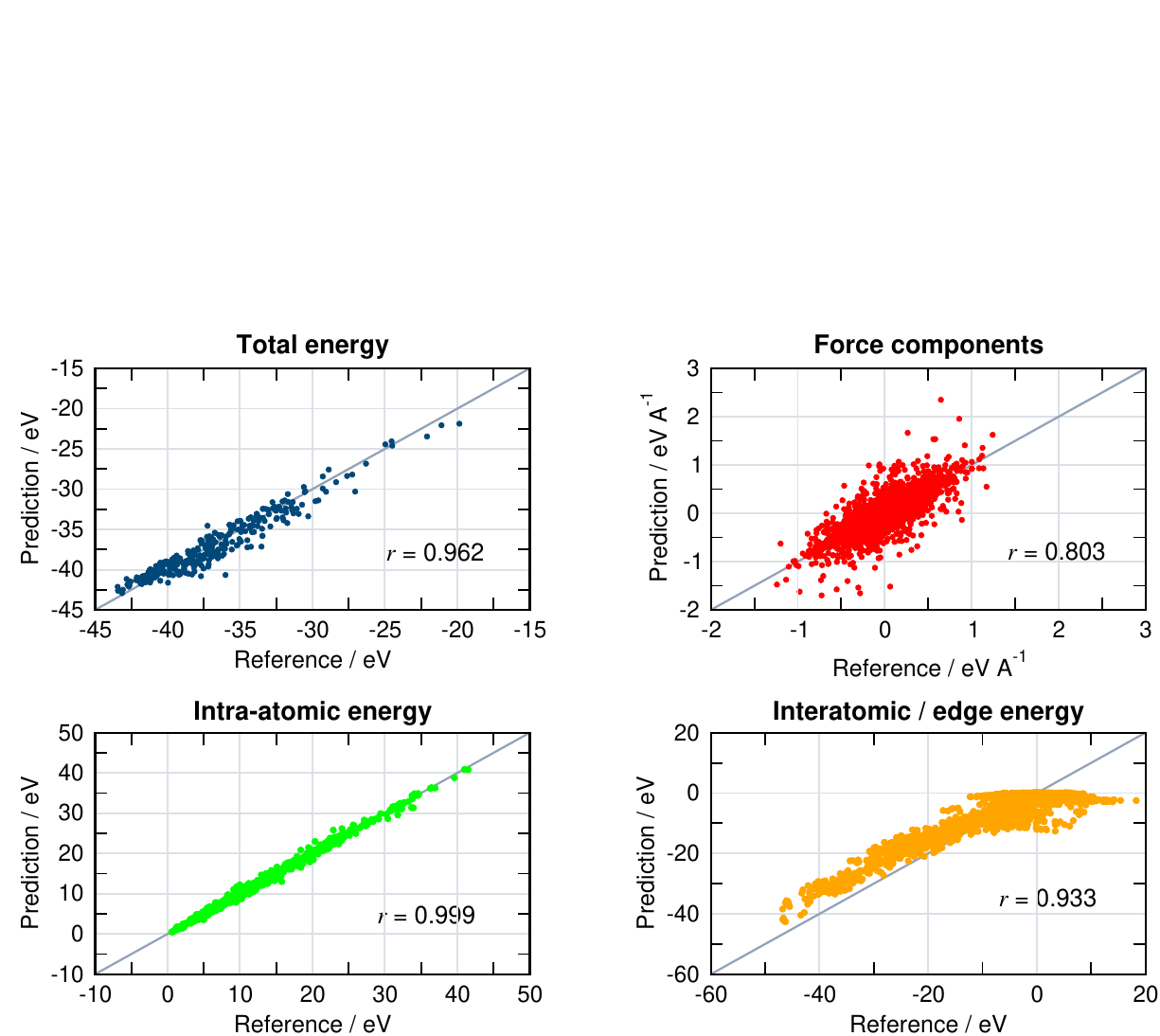}
  \caption{\label{FigS3}
  Parity plots for the model trained with an increased cutoff radius of 5.5 Å. The model is otherwise analogous to IQA-4-9-$\lambda$0.1. For the combined 4-9 atom validation and test data, the correlation coefficients were 0.962 for total energy, 0.803 for force components, 0.999 for $E_{\mathrm{intra}}$, and 0.933 for $E_{\mathrm{inter}}/D_{ij}^{\mathrm{ML}}$. Extending the cutoff from 5.0 to 5.5 Å did not monotonically improve the correspondence between $D_{ij}^{\mathrm{ML}}$ and IQA $E_{\mathrm{inter}}$. This result suggests that the cutoff radius changes not only the presence or absence of weak long-range pairs, but also the entire neighbor graph and the resulting latent energy allocation.}
\end{center}
\end{figure}

\clearpage
\section{Reaction-state-resolved molecular example}
\begin{figure}[h]
\begin{center}
  \includegraphics[width=12cm]{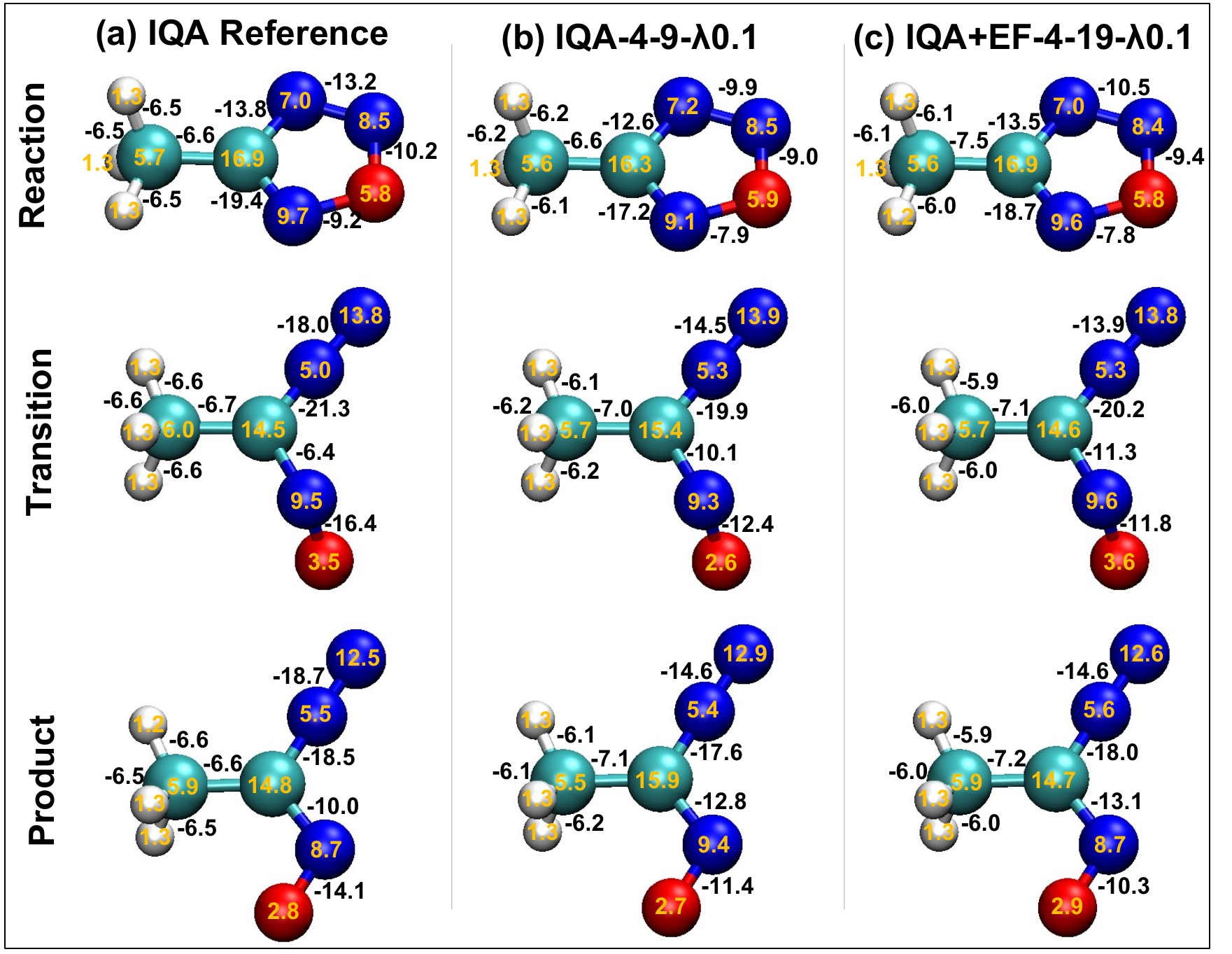}
  \caption{\label{FigS4}
  Molecular-level comparison of the IQA reference decomposition and the corresponding E3D-IQA predictions along a representative Reaction, Transition, and Product sequence from the 4-9 atom training subset. Columns show the IQA reference, IQA-4-9-$\lambda$0.1, and IQA+EF-4-19-$\lambda$0.1. White, cyan, blue, and red spheres represent H, C, N, and O atoms, respectively. Yellow labels denote $E_{\mathrm{intra}}$, and black labels denote IQA $E_{\mathrm{inter}}$ or model-predicted latent pair energy $D_{ij}^{\mathrm{ML}}$. The unit is eV. The learned intra-atomic contribution follows the IQA reference closely in all three states. In the Reaction structure, the main C-H, C-C, C-N, and N-O pair-energy pattern is largely retained. In the Transition and Product structures, larger deviations appear for N/O-centered rearranging pairs, including N-N, N-O, and C-N terms, indicating that the remaining pairwise errors are associated with the sign and magnitude allocation of specific interatomic terms.}
\end{center}
\end{figure}

\clearpage
\section{Element- and pair-resolved parity diagnostics}
To complement the species- and pair-resolved MAEs in Figure 5 of the main text, the decomposition-energy parity plots in Figure 3(b) were separated by element and atom-pair type. This representation shows the reference-energy range, systematic bias, sign mismatch, and outlier structure within each chemical class, thereby helping to identify which classes contribute to the aggregate decomposition errors.

Figure S3 shows that predictions for H, C, N, and O remain concentrated near the identity line over their sampled energy ranges. Thus, the effect of intra-atomic supervision is retained across the element-specific environments represented in the evaluation data.

\begin{figure}[h]
\begin{center}
  \includegraphics[width=12.8cm]{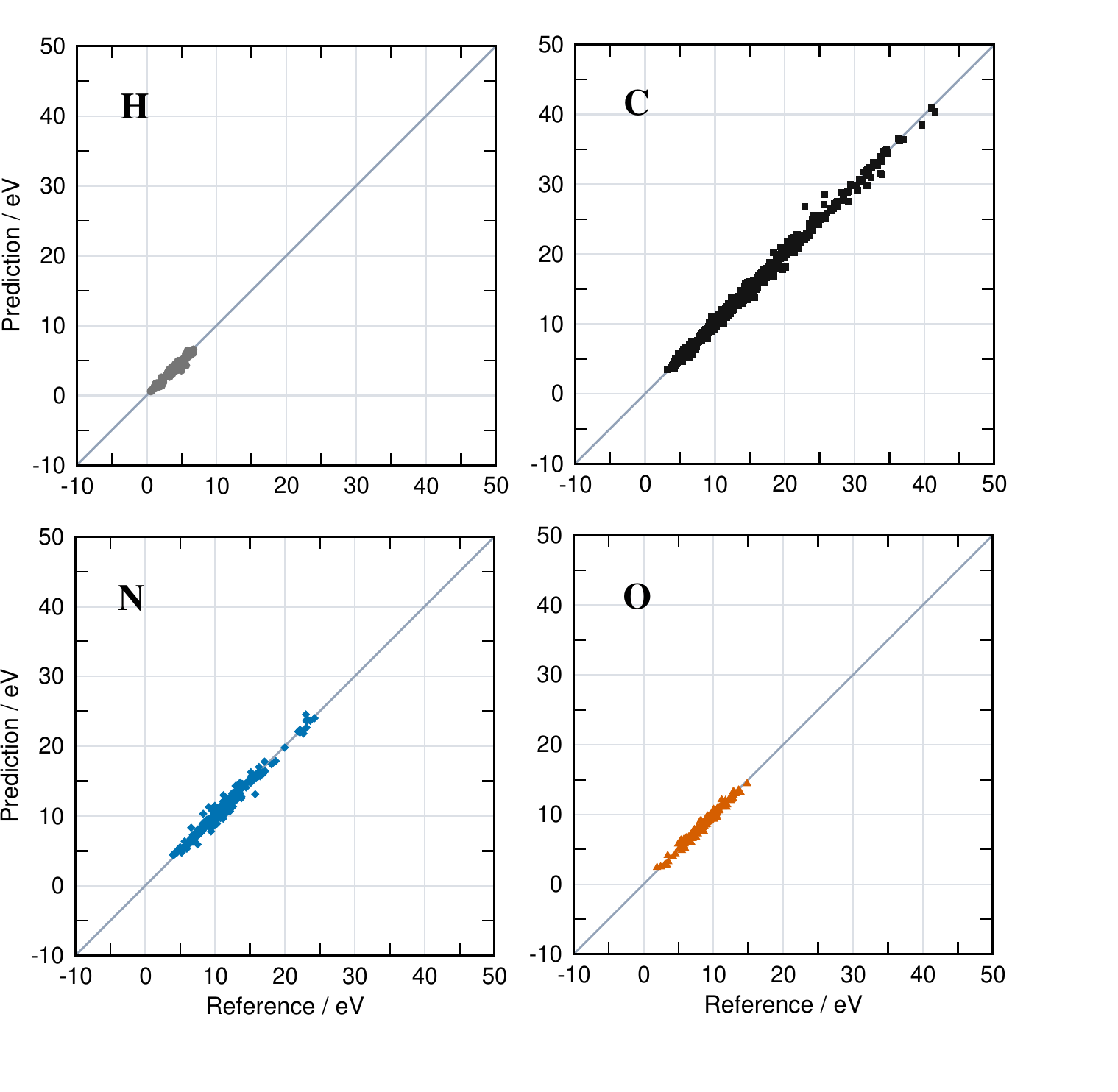}
  \caption{\label{FigS_element_parity}
  Element-resolved parity plots of the predicted node energy $E_{\mathrm{intra}}^{\mathrm{ML}}$ against the IQA $E_{\mathrm{intra}}$ reference for IQA-4-9-$\lambda$0.1 on the combined 4-9 atom validation and test data. The H, C, N, and O panels use common axes and include the identity line.}
\end{center}
\end{figure}

\clearpage
Figure S4 reveals pair-dependent error modes that are obscured by the aggregate $E_{\mathrm{inter}}$ MAE. Strongly negative interactions can retain a monotonic relationship while exhibiting magnitude bias, whereas weak and positive reference interactions are frequently compressed toward zero or assigned negative latent edge energies. The broad C-C reference-energy range and the failure to reproduce positive, destabilizing O-O terms illustrate these distinct error modes.

\begin{figure}[h]
\begin{center}
  \includegraphics[height=0.66\textheight,keepaspectratio]{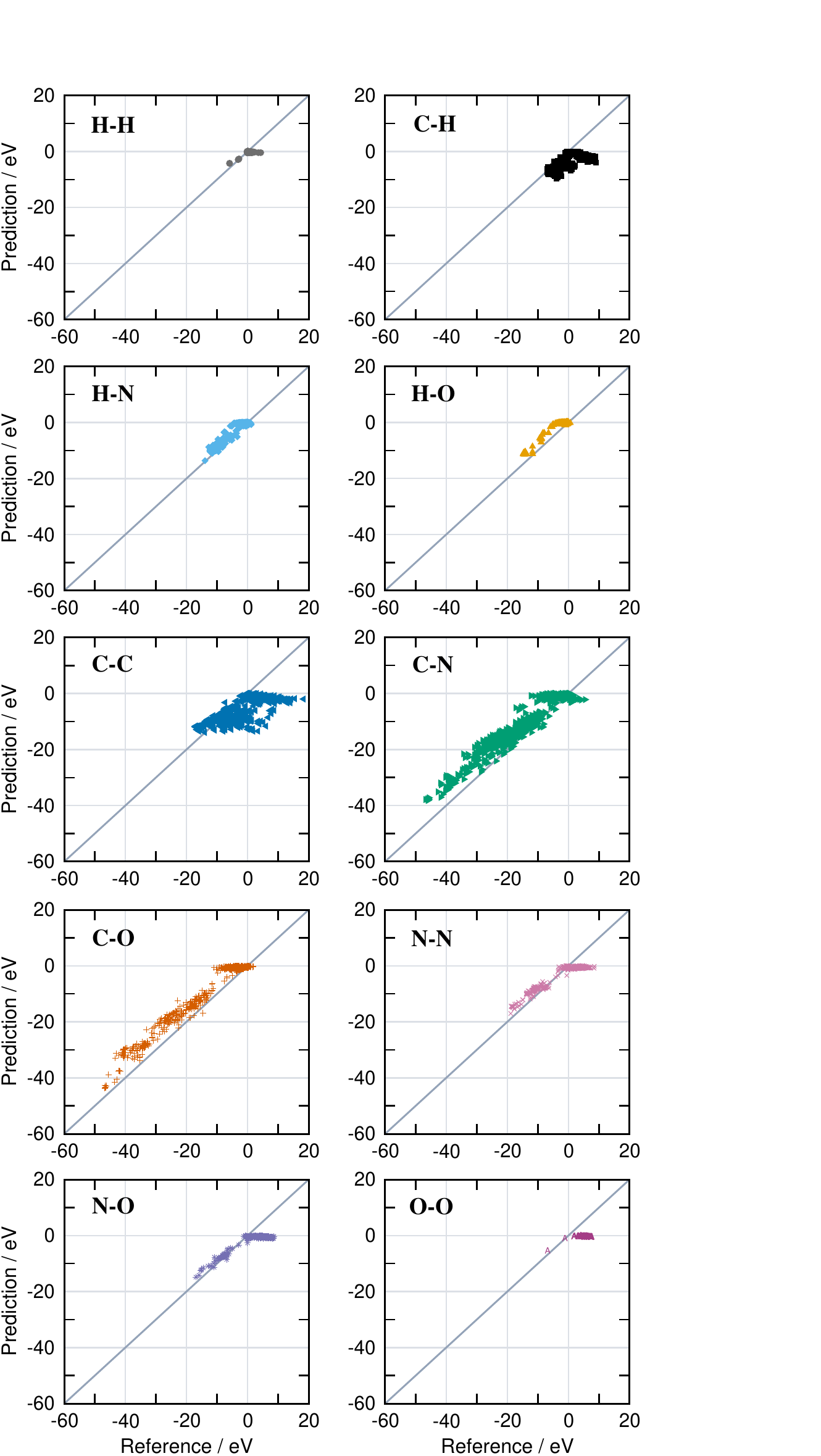}
  \caption{\label{FigS_pair_parity}
  Pair-resolved parity plots of the latent edge energy $D_{ij}^{\mathrm{ML}}$ against the IQA $E_{\mathrm{inter}}$ reference for IQA-4-9-$\lambda$0.1 on the combined 4-9 atom validation and test data. All panels use common axes and include the identity line. Each panel contains all corresponding atom pairs connected by model edges within the 5.0 \AA{} cutoff, including both bonded and nonbonded pairs.}
\end{center}
\end{figure}

\clearpage

%% file: Manuscript.bib
@misc{edamadaka_2025,
      title={Universally Converging Representations of Matter Across Scientific Foundation Models}, 
      author = {Edamadaka, Sathya and Yang, Soojung and Li, Ju and G{\'o}mez-Bombarelli, Rafael},
      year={2025},
      eprint={2512.03750},
      archivePrefix={arXiv},
      primaryClass={cs.LG},
      url={https://arxiv.org/abs/2512.03750}, 
}

@article{Schreiner_2022,
  author    = {Schreiner, Mathias and Bhowmik, Arghya and Vegge, Tejs and Busk, Jonas and Winther, Ole},
  title     = {Transition1x - a dataset for building generalizable reactive machine learning potentials},
  journal   = {Sci. Data.},
  year      = {2022},
  volume    = {9},
  number    = {1},
  pages     = {779},
  month     = {dec},
  doi       = {10.1038/s41597-022-01870-w},
  url       = {https://doi.org/10.1038/s41597-022-01870-w}
}

@article{Maddipati_2026ChemXDyn,
  author  = {Maddipati, Raj and Boddapati, Dhruthi and Arunan, Elangannan and Motamarri, Phani and Aditya, Konduri},
  title   = {{ChemXDyn}: Dynamics-Informed Species and Reaction Detection Methodology from Atomistic Simulations},
  journal = {J. Chem. Theory Comput.},
  year    = {2026},
  volume  = {22},
  number  = {9},
  pages   = {4247--4258},
  doi     = {10.1021/acs.jctc.6c00242}
}

@article{Zhu_2025ReaxANA,
  author  = {Zhu, Hong and Chen, Xin and Gao, Jiali},
  title   = {{ReaxANA}: Analysis of Reactive Dynamics Trajectories for Reaction Network Generation},
  journal = {J. Chem. Inf. Model.},
  year    = {2025},
  volume  = {65},
  number  = {16},
  pages   = {8549--8562},
  doi     = {10.1021/acs.jcim.5c00521}
}

@article{Zeng_2020ReacNetGenerator,
  author  = {Zeng, Jinzhe and Cao, Liqun and Chin, Chih-Hao and Ren, Haisheng and Zhang, John Z. H. and Zhu, Tong},
  title   = {{ReacNetGenerator}: an automatic reaction network generator for reactive molecular dynamics simulations},
  journal = {Physical Chemistry Chemical Physics},
  year    = {2020},
  volume  = {22},
  number  = {2},
  pages   = {683--691},
  doi     = {10.1039/C9CP05091D}
}

@article{Dontgen_2015ChemTraYzer,
  author  = {D{\"o}ntgen, Malte and Przybylski-Freund, Marie-Dominique and Kr{\"o}ger, Leif C. and Kopp, Wassja A. and Ismail, Ahmed E. and Leonhard, Kai},
  title   = {Automated Discovery of Reaction Pathways, Rate Constants, and Transition States Using Reactive Molecular Dynamics Simulations},
  journal = {J. Chem. Theory Comput.},
  year    = {2015},
  volume  = {11},
  number  = {6},
  pages   = {2517--2524},
  doi     = {10.1021/acs.jctc.5b00201}
}

@misc{Keith_2019,
  author       = {Keith, Todd A.},
  title        = {{AIMAll} ({Version} 19.10.12)},
  publisher    = {TK Gristmill Software},
  address      = {Overland Park, KS, USA},
  year         = {2019},
  howpublished = {\url{https://aim.tkgristmill.com/}}
}

@article{Smith_2020,
  author  = {Smith, Daniel G. A. and others},
  title   = {Psi4 1.4: Open-source software for high-throughput quantum chemistry},
  journal = {J. Chem. Phys.},
  volume  = {152},
  number  = {18},
  pages   = {184108},
  year    = {2020},
  doi     = {10.1063/5.0006002}
}

@article{Gallegos_2022,
    author = {Gallegos, Miguel and Guevara-Vela, Jos{\'e} Manuel and Pend{\'a}s, {\'A}ngel Mart{\'i}n},
    title = {NNAIMQ: A neural network model for predicting QTAIM charges},
    journal = {J. Chem. Phys.},
    volume = {156},
    number = {1},
    pages = {014112},
    year = {2022},
    month = {01},
    doi = {10.1063/5.0076896},
}

@article{Takamoto_2022,
title = {TeaNet: Universal neural network interatomic potential inspired by iterative electronic relaxations},
journal = {Comput. Mater. Sci.},
volume = {207},
pages = {111280},
year = {2022},
issn = {0927-0256},
doi = {https://doi.org/10.1016/j.commatsci.2022.111280},
author = {So Takamoto and Satoshi Izumi and Ju Li},
}

@article{Anstine_2025,
    author = {Anstine, Dylan M. and Zubatyuk, Roman and Isayev, Olexandr},
    title = {AIMNet2: a neural network potential to meet your neutral, charged, organic, and elemental-organic needs},
    journal = {Chem. Sci.},
    volume = {16},
    number = {23},
    pages = {10228-10244},
    year = {2025},
    month = {06},
    doi = {10.1039/d4sc08572h},
}

@Article{Guevara_2020,
author = {Guevara-Vela, Jos{\'e} Manuel and Francisco, Evelio and Rocha-Rinza, Tom{\'a}s and Mart{\'i}n Pend{\'a}s, {\'A}ngel},
TITLE = {Interacting Quantum Atoms—A Review},
JOURNAL = {Molecules},
VOLUME = {25},
YEAR = {2020},
NUMBER = {17},
ARTICLE-NUMBER = {4028},
DOI = {10.3390/molecules25174028}
}

@article{Blanco_2005,
    author = {Blanco, M. A. and Martín Pendás, A. and Francisco, E.},
    title = {Interacting Quantum Atoms: A Correlated Energy
Decomposition Scheme Based on the Quantum Theory
of Atoms in Molecules},
    journal = {J. Chem. Theory Comput.},
    volume = {1},
    number = {6},
    pages = {1096-1109},
    year = {2005},
    month = {08},
    doi = {10.1021/ct0501093},
}

@article{Hattori_2026,
  author    = {Hattori, Shinnosuke and Shimamura, Kohei and Nomura, Ken-ichi and Nakano, Aiichiro and Kalia, Rajiv K. and Vashishta, Priya},
  title     = {Chemical intuition on bond-dissociation energies as an emergent ability of universal machine-learning interatomic potentials},
  journal   = {Nat. Commun.},
  year      = {2026},
  month     = {jul},
  issn      = {2041-1723},
  doi       = {10.1038/s41467-026-74919-8},
}

@article{Nomura_2025,
    author = {Nomura, Ken-ichi and Hattori, Shinnosuke and Ohmura, Satoshi and Kanemasu, Ikumi and Shimamura, Kohei and Dasgupta, Nabankur and Nakano, Aiichiro and Kalia, Rajiv K. and Vashishta, Priya},
    title = {Allegro-FM:
Toward an Equivariant Foundation Model
for Exascale Molecular Dynamics Simulations},
    journal = {J. Phys. Chem. Lett.},
    volume = {16},
    number = {25},
    pages = {6637-6644},
    year = {2025},
    month = {06},
    doi = {10.1021/acs.jpclett.5c00605},
}

@article{Batatia_2025,
    author = {Batatia, Ilyes and Benner, Philipp and Chiang, Yuan and Elena, Alin M. and Kovács, Dávid P. and Riebesell, Janosh and Advincula, Xavier R. and Asta, Mark and Avaylon, Matthew and Baldwin, William J. and Berger, Fabian and Bernstein, Noam and Bhowmik, Arghya and Bigi, Filippo and Blau, Samuel M. and Cărare, Vlad and Ceriotti, Michele and Chong, Sanggyu and Darby, James P. and De, Sandip and Della Pia, Flaviano and Deringer, Volker L. and Elijošius, Rokas and El-Machachi, Zakariya and Fako, Edvin and Falcioni, Fabio and Ferrari, Andrea C. and Gardner, John L. A. and Gawkowski, Mikołaj J. and Genreith-Schriever, Annalena and George, Janine and Goodall, Rhys E. A. and Grandel, Jonas and Grey, Clare P. and Grigorev, Petr and Han, Shuang and Handley, Will and Heenen, Hendrik H. and Hermansson, Kersti and Ho, Cheuk Hin and Hofmann, Stephan and Holm, Christian and Jaafar, Jad and Jakob, Konstantin S. and Jung, Hyunwook and Kapil, Venkat and Kaplan, Aaron D. and Karimitari, Nima and Kermode, James R. and Kourtis, Panagiotis and Kroupa, Namu and Kullgren, Jolla and Kuner, Matthew C. and Kuryla, Domantas and Liepuoniute, Guoda and Lin, Chen and Margraf, Johannes T. and Magdău, Ioan-Bogdan and Michaelides, Angelos and Moore, J. Harry and Naik, Aakash A. and Niblett, Samuel P. and Norwood, Sam Walton and O’Neill, Niamh and Ortner, Christoph and Persson, Kristin A. and Reuter, Karsten and Rosen, Andrew S. and Rosset, Louise A. M. and Schaaf, Lars L. and Schran, Christoph and Shi, Benjamin X. and Sivonxay, Eric and Stenczel, Tamás K. and Sutton, Christopher and Svahn, Viktor and Swinburne, Thomas D. and Tilly, Jules and van der Oord, Cas and Vargas, Santiago and Varga-Umbrich, Eszter and Vegge, Tejs and Vondrák, Martin and Wang, Yangshuai and Witt, William C. and Wolf, Thomas and Zills, Fabian and Csányi, Gábor},
    title = {A foundation model for atomistic materials chemistry},
    journal = {J. Chem. Phys.},
    volume = {163},
    number = {18},
    pages = {184110},
    year = {2025},
    month = {11},
    doi = {10.1063/5.0297006},
}

@misc{Kingma_2017,
      title={Adam: A Method for Stochastic Optimization}, 
      author={Diederik P. Kingma and Jimmy Ba},
      year={2017},
      eprint={1412.6980},
      archivePrefix={arXiv},
      primaryClass={cs.LG}
}

@article{Elfwing_2017,
      title={Sigmoid-Weighted Linear Units for Neural Network Function Approximation in Reinforcement Learning},
      author={Stefan Elfwing and Eiji Uchibe and Kenji Doya},
      journal={Neural Netw.},
      year={2018},
      volume={107},
      pages={3--11},
      doi={10.1016/j.neunet.2017.12.012}
}

@misc{Batatia_2023,
      title={MACE: Higher Order Equivariant Message Passing Neural Networks for Fast and Accurate Force Fields}, 
      author={Ilyes Batatia and Dávid Péter Kovács and Gregor N. C. Simm and Christoph Ortner and Gábor Csányi},
      year={2023},
      eprint={2206.07697},
      archivePrefix={arXiv},
      primaryClass={stat.ML}
}

@Article{Batzner_2022,
author={Batzner, Simon
and Musaelian, Albert
and Sun, Lixin
and Geiger, Mario
and Mailoa, Jonathan P.
and Kornbluth, Mordechai
and Molinari, Nicola
and Smidt, Tess E.
and Kozinsky, Boris},
title={E(3)-equivariant graph neural networks for data-efficient and accurate interatomic potentials},
journal={Nat. Commun.},
year={2022},
month={May},
day={04},
volume={13},
number={1},
pages={2453},
doi={10.1038/s41467-022-29939-5}
}

@Article{Musaelian_2023,
author={Musaelian, Albert
and Batzner, Simon
and Johansson, Anders
and Sun, Lixin
and Owen, Cameron J.
and Kornbluth, Mordechai
and Kozinsky, Boris},
title={Learning local equivariant representations for large-scale atomistic dynamics},
journal={Nat. Commun.},
year={2023},
month={Feb},
day={03},
volume={14},
number={1},
pages={579},
doi={10.1038/s41467-023-36329-y}
}

@article{Bader_1985,
Author = {Bader, R. F. W. },
Title = {Atoms in molecules},
Journal = {Acc. Chem. Res.},
Year = {1985},
Volume = {18},
Number = {1},
Pages = {9-15},
DOI = {10.1021/ar00109a003}
}

@article{Behler_2007,
  title = {Generalized Neural-Network Representation of High-Dimensional Potential-Energy Surfaces},
  author = {Behler, J{\"o}rg and Parrinello, Michele},
  journal = {Phys. Rev. Lett.},
  volume = {98},
  issue = {14},
  pages = {146401},
  numpages = {4},
  year = {2007},
  doi = {10.1103/PhysRevLett.98.146401}
}

@article{Musil_2021,
  author = {Musil, Felix and Grisafi, Andrea and Bart{\'o}k, Albert P. and Ortner, Christoph and Cs{\'a}nyi, G{\'a}bor and Ceriotti, Michele},
  title = {Physics-Inspired Structural Representations for Molecules and Materials},
  journal = {Chem. Rev.},
  year = {2021},
  volume = {121},
  number = {16},
  pages = {9759--9815},
  doi = {10.1021/acs.chemrev.1c00021}
}

@article{Schutt_2017,
  author = {Sch{\"u}tt, Kristof T. and Arbabzadah, Farhad and Chmiela, Stefan and M{\"u}ller, Klaus R. and Tkatchenko, Alexandre},
  title = {Quantum-Chemical Insights from Deep Tensor Neural Networks},
  journal = {Nat. Commun.},
  year = {2017},
  volume = {8},
  pages = {13890},
  doi = {10.1038/ncomms13890}
}

@article{Oviedo_2022,
  author = {Oviedo, Felipe and Ferres, Juan Lavista and Buonassisi, Tonio and Butler, Keith T.},
  title = {Interpretable and Explainable Machine Learning for Materials Science and Chemistry},
  journal = {Acc. Mater. Res.},
  year = {2022},
  volume = {3},
  number = {6},
  pages = {597--607},
  doi = {10.1021/accountsmr.1c00244}
}

@article{Kjeldal_2023,
  author = {Kjeldal, Frederik {\O}. and Eriksen, Janus J.},
  title = {Decomposing Chemical Space: Applications to the Machine Learning of Atomic Energies},
  journal = {J. Chem. Theory Comput.},
  year = {2023},
  volume = {19},
  number = {7},
  pages = {2029--2038},
  doi = {10.1021/acs.jctc.2c01290}
}

@article{Chun_2025,
  author = {Chun, Hoje and Hong, Minjoon and Noh, Seung Hyo and Han, Byungchan},
  title = {Learning Pairwise Interaction for Extrapolative and Interpretable Machine Learning Interatomic Potentials with Physics-Informed Neural Network},
  journal = {J. Chem. Theory Comput.},
  year = {2025},
  volume = {21},
  number = {8},
  pages = {4030--4039},
  doi = {10.1021/acs.jctc.5c00090}
}

@article{Symons_2019,
  author = {Symons, Benjamin C. B. and Williamson, David J. and Brooks, Chloe M. and Wilson, Antonia L. and Popelier, Paul L. A.},
  title = {Does the Intra-Atomic Deformation Energy of Interacting Quantum Atoms Represent Steric Energy?},
  journal = {ChemistryOpen},
  year = {2019},
  volume = {8},
  number = {5},
  pages = {560--570},
  doi = {10.1002/open.201800275}
}

@article{JaraCortes_2021,
  author = {Jara-Cort{\'e}s, Jes{\'u}s and Leal-S{\'a}nchez, Edith and Francisco, Evelio and P{\'e}rez-Pimienta, Jos{\'e} A. and Mart{\'i}n Pend{\'a}s, {\'A}ngel and Hern{\'a}ndez-Trujillo, Jes{\'u}s},
  title = {Implementation of the Interacting Quantum Atom Energy Decomposition Using the {CASPT2} Method},
  journal = {Phys. Chem. Chem. Phys.},
  year = {2021},
  volume = {23},
  number = {48},
  pages = {27508--27519},
  doi = {10.1039/D1CP02837E}
}

@article{Maxwell_2016,
  author = {Maxwell, Peter and Mart{\'i}n Pend{\'a}s, {\'A}ngel and Popelier, Paul L. A.},
  title = {Extension of the Interacting Quantum Atoms ({IQA}) Approach to {B3LYP} Level Density Functional Theory ({DFT})},
  journal = {Phys. Chem. Chem. Phys.},
  year = {2016},
  volume = {18},
  pages = {20986--21000},
  doi = {10.1039/C5CP07021J}
}

@article{Becke_1993,
  author = {Becke, Axel D.},
  title = {Density-Functional Thermochemistry. {III}. The Role of Exact Exchange},
  journal = {J. Chem. Phys.},
  year = {1993},
  volume = {98},
  number = {7},
  pages = {5648--5652},
  doi = {10.1063/1.464913}
}

@article{Lee_1988,
  author = {Lee, Chengteh and Yang, Weitao and Parr, Robert G.},
  title = {Development of the {Colle--Salvetti} Correlation-Energy Formula into a Functional of the Electron Density},
  journal = {Phys. Rev. B},
  year = {1988},
  volume = {37},
  number = {2},
  pages = {785--789},
  doi = {10.1103/PhysRevB.37.785}
}

@article{Weigend_2005,
  author = {Weigend, Florian and Ahlrichs, Reinhart},
  title = {Balanced Basis Sets of Split Valence, Triple Zeta Valence and Quadruple Zeta Valence Quality for {H} to {Rn}: Design and Assessment of Accuracy},
  journal = {Phys. Chem. Chem. Phys.},
  year = {2005},
  volume = {7},
  pages = {3297--3305},
  doi = {10.1039/B508541A}
}

@article{Rappoport_2010,
  author = {Rappoport, Dmitrij and Furche, Filipp},
  title = {Property-Optimized Gaussian Basis Sets for Molecular Response Calculations},
  journal = {J. Chem. Phys.},
  year = {2010},
  volume = {133},
  number = {13},
  pages = {134105},
  doi = {10.1063/1.3484283}
}
